\documentclass[twocolumn,showpacs,prb]{revtex4}
\usepackage{graphicx,amsmath,dsfont}

\newcommand{\rr}{{\boldsymbol r}}
\newcommand{\rs}{{{\boldsymbol r}'}}
\newcommand{\rrs}{{{\boldsymbol r}{\boldsymbol r}'}}
\newcommand{\rsr}{{{\boldsymbol r}'{\boldsymbol r}}}

\newcommand{\bs}{\boldsymbol}
\newcommand{\pair}{{({\boldsymbol r},{\boldsymbol r}')}}
\newcommand{\T}{\mathcal{T}}
\newcommand{\drr}{{\delta{\boldsymbol r}}}
\newcommand{\dx}{{\delta x}}
\newcommand{\dy}{{\delta y}}
\newcommand{\kk}{{\boldsymbol k}}

\begin{document}

\title{Classification of spin liquids on the square lattice with strong spin-orbit coupling}
\author{Johannes Reuther, Shu-Ping Lee, and Jason Alicea}
\affiliation{Department of Physics and Institute for Quantum Information and Matter, California Institute of Technology, Pasadena, California 91125, USA}

%%%%%%%%%%%%%%%%%%%%%%%%%%%%%%%%%%%%%%%%%%%%%%%%%%%%%%%%%%%%%%
% ABSTRACT
%%%%%%%%%%%%%%%%%%%%%%%%%%%%%%%%%%%%%%%%%%%%%%%%%%%%%%%%%%%%%%

\begin{abstract}
Spin liquids represent exotic types of quantum matter that evade conventional symmetry-breaking order even at zero temperature.  
Exhaustive classifications of spin liquids have been carried out in several systems, particularly in the presence of full SU(2) spin-rotation symmetry.  Real magnetic compounds, however, generically break SU(2) spin symmetry as a result of spin-orbit coupling---which in many materials provides an `order one' effect.  We generalize previous works by using the projective symmetry group method to classify $\mathds{Z}_2$ spin liquids on the square lattice when SU(2) spin symmetry is maximally lifted. We find that, counterintuitively, the lifting of spin symmetry actually results in vastly more spin liquid phases compared to SU(2)-invariant systems.  A generic feature of the SU(2)-broken case is that the spinons naturally undergo $p+ip$ pairing; consequently, many of these $\mathds{Z}_2$ spin liquids feature a topologically nontrivial spinon band structure supporting gapless Majorana edge states.   
We study in detail several spin-liquid phases with varying numbers of gapless edge states and discuss their topological protection. The edge states are often protected by a combination of time reversal and lattice symmetries and hence resemble recently proposed topological crystalline superconductors.
\end{abstract}


\maketitle

%%%%%%%%%%%%%%%%%%%%%%%%%%%%%%%%%%%%%%%%%%%%%%%%%%%%%%%%%%%%%%
% I. INTRODUCTION
%%%%%%%%%%%%%%%%%%%%%%%%%%%%%%%%%%%%%%%%%%%%%%%%%%%%%%%%%%%%%%

\section{Introduction}
When cooled down to low temperatures, materials normally develop a variety of different orders such as magnetism or superconductivity. However, sufficiently severe quantum fluctuations may prevent the formation of any type of symmetry-breaking order---even at absolute zero---leading to a liquid-like ground state. In the case of spinful quantum systems, such ``spin liquid" phases\cite{balents10,anderson73,fazekas74} have attracted enormous interest in recent years. Spin liquids come in many flavors and can support either gapless or fully gapped spin excitations; importantly, however, they are always far from featureless.  For instance, gapped spin liquids are not just characterized by the absence of spontaneously broken symmetries, but also by the existence of fractionalized spinon excitations.\cite{read91,wen91} In such states the well-known concept of conventional symmetry breaking is replaced by topological order associated with long-range quantum entanglement.\cite{isakov11}

In recent years, mounting experimental evidence suggests that spin liquids could be realized in strongly frustrated magnetic materials, e.g., in the Kagome lattice compound Herbertsmithite.\cite{helton07,mendels07,han12} Furthermore, novel classes of magnetic compounds with strong spin-orbit coupling such as iridium oxides have opened a new arena in this field of research. Generally, spin-orbit coupling breaks SU(2) spin-rotation symmetry, which may generate additional frustration effects leading to exotic magnetic or non-magnetic spin phases.\cite{chaloupka10, jackeli09,pesin10,reuther12,ruegg12,kimchi14} Prominent examples are the iridate compounds Na$_2$IrO$_3$ and Li$_2$IrO$_3$ (see, e.g., Refs.~\onlinecite{choi12,singh12,liu11}) where strong spin-orbit coupling has been proposed to realize the Kitaev spin model, which is known to harbor a spin liquid.\cite{kitaev06} In contrast to many other more conventional systems, where spin-orbit coupling constitutes a small relativistic perturbation that may at first pass be ignored, a realistic description of such materials needs to take into account SU(2) spin-symmetry breaking. The systematic investigation of the effects of strong spin-orbit coupling on spin-liquid phases is the purpose of this paper.

Since spin liquids are inherently strongly coupled quantum systems, their theoretical investigation represents a major challenge in condensed matter physics. Given a generic spin Hamiltonian, the unambiguous identification of a spin liquid ground state is not possible by presently available analytical means. Numerical approaches on the other hand, while certainly invaluable, tend to be limited in their applicability.  Instead of trying to solve a spin model, one can alternatively pursue a more `universal' strategy: identifying all allowed spin liquid states based on a system's symmetries, which can yield valuable insights in the ultimate quest for experimental realization.  

The projective symmetry-group (PSG) method first proposed by Wen\cite{wen02,wen02_2} represents one powerful way of systematically classifying possible spin liquids for a given quantum spin system. This approach makes use of a fermionic parton\cite{abrikosov65} (spinon) description for spin operators and imposes a constraint on the total number of fermions on each site (extensions of the method have also been developed for Schwinger bosons\cite{read91,sachdev92,wang06}). Depending on the specific symmetries present, the resulting fermionic Hamiltonian can be mean-field decoupled in various ways, leading to a quadratic model that can be treated analytically.  After a Gutzwiller projection onto the physically occupied subspace, one obtains a physical trial spin wavefunction. While in general a mean-field ansatz partially breaks the local SU(2) gauge freedom of the fermionic parton fields, the remaining gauge group determines the nature of gauge fluctuations around a mean-field solution.  The effects of gauge fluctuations, possibly destabilizing a spin liquid state, can be subtle.  For example, if the gauge structure is U(1), it has been argued that gauge fluctuations can drive a topological state into a conventional magnetically ordered phase.\cite{ran09} On the other hand, there are also cases where U(1) spin liquids in two dimensions are found to be stable.\cite{hermele04}  To avoid subtleties associated with gauge fluctuations, we restrict our analysis to the simplest case of $\mathds{Z}_2$ spin liquids, where all gauge excitations are gapped. In this situation a spin-liquid solution faithfully describes a physical quantum spin phase that should be stable beyond mean-field treatment.

So far the PSG method has been applied to various two-dimensional systems such as spins on the square\cite{wen02,chen12, essin13}, triangular\cite{zhou02,wang06,messio13}, honeycomb\cite{lu11,wang10}, and Kagome lattices\cite{wang06,lu11_2,ran07,messio12}. Even three dimensional systems may be treated.\cite{kou09,lawler08,burnell09,hwang13} %While most of these works consider spin-{\it isotropic} models, systematic studies that explore the case with broken SU(2) spin symmetry are relatively scarce (for some interesting examples see Refs.~\onlinecite{schaffer13} and~\onlinecite{dodds13}). 
The PSG method typically identifies a large number of $\mathds{Z}_2$ spin liquids. For square-lattice systems preserving SU(2) spin symmetry as well as all spatial symmetries and time reversal, Wen originally identified 272 $\mathds{Z}_2$ phases.\cite{wen02}  We note that Ref.~\onlinecite{essin13} introduced a related classification scheme that also employs projective representations of symmetries. This approach does not rely on a specific parton construction but directly considers the action of projective symmetries on anyonic excitations, leading to a coarser classification. 

Most of the preceding works considered spin-{\it isotropic} models; systematic studies that explore the case with broken SU(2) spin symmetry are, by contrast,  relatively scarce (for some interesting examples see Refs.~\onlinecite{schaffer13} and~\onlinecite{dodds13}). In this work we generalize previous studies and classify $\mathds{Z}_2$ spin liquids on the square lattice when SU(2) spin symmetry is lifted strongly by spin-orbit interactions. Naively, one might expect that the reduced number of symmetries also reduces the number of spin liquid solutions. This intuition may be justified by the results of a PSG analysis on the square lattice where the removal of all point-group symmetries obliterates the aforementioned 272 states leaving just 16 phases.\cite{wen02} Here, however, we find quite the opposite: Lifting the SU(2) spin symmetry increases the number of spin liquids from 272 in the spin-isotropic case to $272 + 1488 = 1760$. Most strikingly, the 1488 new states exist {\it solely} because SU(2) symmetry is broken, i.e., they have no analogue in SU(2)-symmetric spin systems. [We note that eight of these 1488 states still fulfill a projective version of the SU(2) spin symmetry and should therefore rather be added to the 272 solutions; see Ref.~\onlinecite{chen12}. For simplicity, however, we  neglect such subtleties in the following and treat continuos spin symmetries non-projectively.]  

The dramatic refinement of the $\mathds{Z}_2$ spin liquid classification upon breaking SU(2) spin symmetry can be understood as follows.  In the absence of continuous spin rotation symmetry spin liquids need to fulfill an {\it additional} symmetry corresponding to invariance under spatial reflection about the two-dimensional lattice plane ($z\rightarrow -z$). This symmetry operation does not transform the positions of the lattice sites but still rotates the spins. While in spin-isotropic systems a spin rotation is trivial, the reflection may act nontrivially in the absence of SU(2) spin symmetry.  This new symmetry condition leads to additional combinations of spin-liquid solutions, effectively increasing their number.  We emphasize that this conclusion is quite general and is expected to refine the PSG classification in any two-dimensional system with $z\rightarrow -z$ symmetry.   

The increased richness of spin-liquid solutions has some remarkable consequences. The structure of the mean-field ans\"atze as obtained by the PSG approach generally supports $p+ip$ triplet pairing for the spinons which may, in turn, lead to gapless Majorana edge states.  This type of pairing is not possible in spin-isotropic systems. Solving the mean-field Hamiltonians on a cylinder, we indeed identify various $\mathds{Z}_2$ spin-liquid phases with a gapped, topologically nontrivial bulk supporting gapless edge states. Even though spin-liquid solutions differ in the number of boundary modes and may have elaborate edge-state structures, we find that their topological protection can be traced back to simple $\mathds{Z}_2$ topological indices based on time-reversal invariance. Furthermore, we identify various spin liquids where additional lattice symmetries are needed to protect the gaplessness of the boundary modes. This demonstrates that spinon band structures resembling the recently proposed topological crystalline superconductors\cite{fu11,teo13} may be realized in magnetic compounds and accessed theoretically using the PSG method. Such an interplay between topology and symmetry has been described in the general framework of ``symmetry enriched topological phases"\cite{chen13,mesaros13,lu13}: If a topologically ordered system has additional global symmetries, new classes of distinct quantum phases can emerge. In our case, many spin liquids have the remarkable property that a non-trivial topology is implemented in {\it two} different ways, via the existence of deconfined spinon excitations and by a symmetry-protected topological spinon band structure, similar to Refs.~\onlinecite{young08,rachel10,krempa10,pesin10,swingle11,ruegg12,cho12}.

We structure the remainder of the paper as follows. In Sec.~\ref{su2} we briefly revisit the case of spin-isotropic systems and summarize some fundamentals of the PSG approach. A classification of spin liquids when SU(2) spin symmetry is broken down to U(1) is discussed in Sec.~\ref{u1}. We then turn in Sec.~\ref{no_symmetry} to a complete classification of spin liquids when spin symmetry is maximally lifted and discuss the modifications of the PSG method required for such a generalization. Afterwards, specific examples for spin liquid solutions are presented in Sec.~\ref{examples}. There we particularly focus on the description of boundary modes and their topological protection. Section~\ref{discussion} summarizes our main results and discusses future directions.  Two appendices list the explicit implementations of projective symmetries within all classes of $\mathds{Z}_2$ spin liquids.

%%%%%%%%%%%%%%%%%%%%%%%%%%%%%%%%%%%%%%%%%%%%%%%%%%%%%%%%%%%%%%
% II SPIN-ISOTROPIC Z_2 SPIN LIQUIDS
%%%%%%%%%%%%%%%%%%%%%%%%%%%%%%%%%%%%%%%%%%%%%%%%%%%%%%%%%%%%%%

\section{Spin-isotropic $\mathds{Z}_2$ spin liquids}\label{su2}
Before discussing the more complicated situation where SU(2) spin symmetry is lifted, we first revisit the PSG classification of spin liquids on a square lattice in the spin-isotropic case.\cite{wen02} Our starting point is a general spin-1/2 Hamiltonian of the form
\begin{equation}
H=\sum_\pair J_\rrs{\bs S}_\rr\cdot{\bs S}_\rs+\cdots\;,\label{hamiltonian}
\end{equation}
where $\rr$, $\rs$ label the square-lattice sites and $\pair$ denotes pairs of sites. The ellipsis indicates that interactions with more than two spin operators may be present as well. We emphasize that at this point we do not need to further specify the Hamiltonian. The only requirements are that $H$ is spin-isotropic, time-reversal invariant and respects all lattice symmetries. Next, the spin operators are expressed in terms of fermionic parton operators via\cite{abrikosov65}
\begin{equation}
S_\rr^j=\frac{1}{2}f_\rr^\dagger \sigma^j f_\rr\;,\label{fermions}
\end{equation}
where $f_\rr=(f_{\rr\uparrow},f_{\rr\downarrow})^\text{T}$ is a two-component spinor and $f_{\rr\alpha}^\dagger$ creates a fermion of spin $\alpha=\uparrow,\downarrow$ at site $\rr$. Pauli matrices operating in spin space are denoted by $\sigma^j$ ($j=1,2,3$). The fermionic representation in Eq.~(\ref{fermions}) comes along with an artificial enlargement of the Hilbert space. The physical spin-1/2 subspace is singled out by the constraints
\begin{equation}
\sum_\alpha f_{\rr\alpha}^\dagger f_{\rr\alpha}=1\;,\quad f_{\rr\uparrow}f_{\rr\downarrow}=0\;,\label{constraint}
\end{equation}
which only allow for {\it one} fermion on each lattice site. Note that the second constraint in Eq.~(\ref{constraint}) is a consequence of the first. Since the physical spin-1/2 operators are only defined in a subspace of the fermions, it is clear that the fermionic system must exhibit a gauge redundancy. Given the two-component spinor
\begin{equation}
\psi_\rr=(f_{\rr\uparrow},f_{\rr\downarrow}^\dagger)^\text{T}\label{psi_1}
\end{equation}
and a $2\times 2$ SU(2) matrix $W_\rr$ with $W_\rr^\dagger=W_\rr^{-1}$ one can easily check that a transformation $\psi_\rr\rightarrow W_\rr \psi_\rr$ leaves the physical spin operator ${\boldsymbol S}_\rr$ unchanged. Hence, for a fermionic version of the spin Hamiltonian (\ref{hamiltonian}) there is a {\it local} SU(2) gauge freedom corresponding to transformations with {\it site dependent} matrices $W_\rr$.

%%%%%%%%%%%%%%%%%%%%%%%%%%%%%%%%%%%%%%%%%%%%%%%%%%%%%%%%%%%%%%
% II.A. MEAN-FIELD DECOUPLING OF THE HAMILTONIAN
%%%%%%%%%%%%%%%%%%%%%%%%%%%%%%%%%%%%%%%%%%%%%%%%%%%%%%%%%%%%%%

\subsection{Mean-field decoupling of the spin Hamiltonian}
The Hamiltonian in Eq.~(\ref{hamiltonian}) in the fermionic representation of Eq.~(\ref{fermions}) is now treated within a mean-field approach. To this end, we rewrite $H$ by performing the most general non-magnetic mean-field decoupling with the mean-field amplitudes $\chi_\rrs=\langle f_{\rr\uparrow}^\dagger f_{\rs\uparrow}\rangle=\langle f_{\rr\downarrow}^\dagger f_{\rs\downarrow}\rangle$ (hopping) and $\eta_\rrs=\langle f_{\rr\uparrow}f_{\rs\downarrow}\rangle$ (pairing). (By ``non-magnetic" we mean that correlators $\langle S_\rr^\mu\rangle=\frac{1}{2}\langle f_\rr^\dagger \sigma^\mu f_\rr\rangle$ vanish.) A mean-field treatment is also performed for the occupancy constraint in Eq.~(\ref{constraint}) by replacing the exact constraint by its ground-state expectation value,
\begin{equation}
\sum_\alpha\langle f_{\rr\alpha}^\dagger f_{\rr\alpha}\rangle=1\;,\quad \langle f_{\rr\uparrow}f_{\rr\downarrow}\rangle=0\;.\label{mf_constraint}
\end{equation} 
These weaker conditions can be enforced by adding on-site terms $\sum_\rr\{a_3(f_\rr^\dagger f_\rr-1)+[(a_1+ia_2)f_{\rr\downarrow}f_{\rr\uparrow}+\text{h.c.}]\}$ to the Hamiltonian, where $a_j$ ($j=1,2,3$) are real Lagrange multipliers. The full spin-isotropic mean-field Hamiltonian then reads
\begin{eqnarray}
H_\text{mf}&=\negthickspace&\sum_\pair\left[\chi_\rsr f_\rr^\dagger f_\rs+\eta_\rrs\left(f_{\rr\uparrow}^\dagger f_{\rs\downarrow}^\dagger-f_{\rr\downarrow}^\dagger f_{\rs\uparrow}^\dagger\right)+\text{h.c.}\right]\notag\\
&&+\sum_\rr\{a_3(f_\rr^\dagger f_\rr-1)+[(a_1+ia_2)f_{\rr\downarrow}f_{\rr\uparrow}+\text{h.c.}]\}\;.\notag\\\label{mf_hamiltonian_f}
\end{eqnarray}
The SU(2) spin symmetry is explicit here which can be seen by performing a spin rotation $f_\rr\rightarrow W_\text{spin} f_\rr$ with a $2\times 2$ SU(2) matrix $W_\text{spin}^\dagger=W_\text{spin}^{-1}$, leaving the mean-field Hamiltonian invariant.

In the following it will be convenient to express $H_\text{mf}$ in terms of the spinors $\psi_\rr$ defined in Eq.~(\ref{psi_1}), leading to
\begin{equation}
H_\text{mf}=\sum_\pair\left(\psi_\rr^\dagger u_\rrs \psi_\rs +\text{h.c.}\right)+\sum_\rr\sum_{j=1}^3 a_j\psi_\rr^\dagger\tau^j\psi_\rr\;,\label{mf_hamiltonian_full}
\end{equation}
where the $2\times 2$ matrix $u_\rrs$ contains the mean-field amplitudes defined above,
\begin{equation}
u_\rrs=
\left(\begin{array}{cc}
\chi_\rrs^* & \eta_\rrs \\
\eta_\rrs^* & -\chi_\rrs
\end{array}\right)\;.\label{u_ij}
\end{equation}
Note that Hermiticity of $H_\text{mf}$ requires $u_\rrs=u_\rsr^\dagger$. In the spin-isotropic case considered in this section, $u_\rrs$ may also be conveniently written as 
\begin{equation}
u_\rrs=i s^0_\rrs\tau^0+\sum_{j=1}^3 s^j_\rrs \tau^j\;,\label{u_ij2}
\end{equation}
with real parameters $s^\mu$ ($\mu=0,1,2,3$). Here and in the rest of the paper $\tau^\mu$ are $2\times 2$ matrices operating in the ``Nambu-space" of the spinor $\psi_\rr$, where $\tau^0=\mathds{1}_{2\times2}$ and $\tau^1$, $\tau^2$, $\tau^3$ are Pauli matrices.

Since $H_\text{mf}$ is quadratic in the fermions it can be easily diagonalized. However, the ground state as well as all other states inevitably violate the exact constraint of Eq.~(\ref{constraint}). To obtain a physical spin state, the ground state needs to be Gutzwiller projected onto the singly occupied subspace. Furthermore, once the mean-field amplitudes $u_\rrs$ and $a_j$ are fixed (i.e., quantum fluctuations in these parameters are ignored), $H_\text{mf}$ breaks the local SU(2) gauge freedom. A transformation $\psi_\rr\rightarrow W_\rr\psi_\rr$ gives rise to new mean-field parameters $u_\rrs\rightarrow W_\rr^\dagger u_\rrs W_\rs$ which are (at least in the generic case) different from the original ones, leading to a different mean-field ground state. However, since gauge transformations leave the physical spin sector invariant, {\it after} projection all different gauge choices result in the same wave function.

It is important to note that a mean-field ansatz cannot {\it completely} break the local SU(2) gauge freedom. Any Hamiltonian $H_\text{mf}$ may still be gauge invariant under a {\it global} SU(2), U(1) or $\mathds{Z}_2$ transformation. In these cases, there exists a subgroup of the local $SU(2)\times SU(2)\times\cdots$ gauge group such that
\begin{equation}
u_\rrs=W_\rr^\dagger u_\rrs W_\rs\;.\label{igg}
\end{equation}
More generally, $u_\rrs$ must {\it at least} be invariant under a global $\mathds{Z}_2$ gauge transformation, since Eq.~(\ref{igg}) is always fulfilled for (site independent) transformations with $W_\rr=\pm\tau^0$. As first noted by Wen\cite{wen91,wen02}, the remaining gauge freedom -- also called the invariant gauge group (IGG) -- determines the structure of gauge fluctuations around a mean-field ansatz and is, hence, an important criterion for the stability of an ansatz beyond mean-field theory. The effect of gauge fluctuations is particularly simple for mean-field ans\"atze with a $\mathds{Z}_2$ IGG. In this case all gauge fluctuations are gapped such that possible low energy spinon excitations of a mean-field solution should be stable against fluctuations. Thus, the fermions introduced above still represent effective quasiparticles of the system even if quantum fluctuation are included, i.e., they directly correspond to deconfined spinon excitations. Such states can therefore be faithfully regarded as stable quantum phases. We hence restrict our investigations in this paper to ans\"atze with a $\mathds{Z}_2$ gauge structure.

%%%%%%%%%%%%%%%%%%%%%%%%%%%%%%%%%%%%%%%%%%%%%%%%%%%%%%%%%%%%%%
% II.B. PROJECTIVE IMPLEMENTATION OF SYMMETRIES
%%%%%%%%%%%%%%%%%%%%%%%%%%%%%%%%%%%%%%%%%%%%%%%%%%%%%%%%%%%%%%

\subsection{Projective implementation of symmetries}
To construct spin liquids within the PSG method, one needs to ensure that after projection the ground states of the mean-field Hamiltonians $H_\text{mf}$ preserve all physical symmetries of the system. The SU(2) spin symmetry is already guaranteed by the form of the Hamiltonian in Eq.~(\ref{mf_hamiltonian_f}). Furthermore, for a square lattice in the $x$-$y$ plane one has to enforce translation symmetries in the $x$ and $y$-directions (in the following called $T_x$ and $T_y$), mirror symmetries about the $x$ and $y$-directions ($P_x$ and $P_y$), mirror symmetry along a diagonal interchanging $x\leftrightarrow y$ ($P_{xy}$)\cite{pxy}, and a symmetry $P_z$ which performs a reflection $z\rightarrow -z$ about the square lattice plane. Using $\rr=(x,y)$ the action of these symmetries on the lattice coordinates is given by
\begin{eqnarray}
&T_x:\;\rr\rightarrow(x+1,y)\qquad
T_y:\;\rr\rightarrow(x,y+1)&\notag\\
&P_x:\;\rr\rightarrow(-x,y)\qquad
P_y:\;\rr\rightarrow(x,-y)&\notag\\
&P_{xy}:\;\rr\rightarrow(y,x)\qquad P_z:\;\rr\rightarrow\rr\;,&\label{define_symmetries}
\end{eqnarray}
where the lattice constant is set to unity. Additionally, we want to enforce time-reversal invariance $\T$. For a spinful model, $\T$ is an antiunitary operator which, when applied twice, reverses the sign of an arbitrary single-particle wave function $|\phi\rangle$, i.e., $\T^2|\phi\rangle=-|\phi\rangle$.  Furthermore, up to a SU(2) gauge transformation, it transforms the spinors $f_\rr$ via $\T f_\rr \T^\dagger =i\sigma^2 f_\rr$. Equivalently, the action on $\psi_\rr$ is given by $\T \psi_\rr \T^\dagger=[(-i\tau^2\psi_\rr)^\text{T}]^\dagger$. To simplify the time-reversal operation we perform an additional gauge transformation $\psi_\rr\rightarrow i\tau^2\psi_\rr$ such that $\T \psi_\rr \T^\dagger=(\psi_\rr^\text{T})^\dagger$.\cite{t}

When applied to a PSG mean-field Hamiltonian, symmetry operations effectively transform the matrices $u_\rrs$. Particularly, time reversal $\T$ acts as
\begin{equation}
\T:\;u_\rrs\rightarrow -u_\rrs\;,
\end{equation}
while the spatial symmetries $\mathcal{S}'=\{T_x,T_y,P_x,P_y,P_{xy}\}$ are implemented via
\begin{equation}
\mathcal{S}':\;u_\rrs\rightarrow u_{\mathcal{S}'(\rr)\mathcal{S}'(\rs)}\;.
\end{equation}
In principle, the inversion symmetry $P_z$ also needs to be considered. While such a symmetry operation does not transform the lattice coordinates, it still rotates the spins. However, since inversion $P_z$ is a subgroup of the SU(2) symmetry, this spin rotation has no effect on a spin-isotropic Hamiltonian such that $P_z$ can be ignored. When we lift the SU(2) spin symmetry in subsequent sections, the action of $P_z$ will play a central role.

In an ordinary fermionic system where the fermions are fundamental particles, lattice symmetries can be enforced by the condition $u_\rrs=u_{\mathcal{S}'(\rr)\mathcal{S}'(\rs)}$. However, in a fermionic version of a spin system with SU(2) gauge freedom, every symmetry operation may be supplemented with an additional gauge transformation. Hence, a spatial symmetry $\mathcal{S}'$ is already fulfilled if there are (site dependent) $2\times 2$ SU(2) matrices $G^{\mathcal{S}'}_\rr$ such that
\begin{equation}
u_\rrs=G^{\mathcal{S}'\dagger}_{\mathcal{S}'(\rr)}u_{\mathcal{S}'(\rr)\mathcal{S}'(\rs)}G^{\mathcal{S}'}_{\mathcal{S}'(\rs)}\;.\label{psg_s}
\end{equation}
Similarly, time-reversal invariance is fulfilled if matrices $G^\T_\rr$ with
\begin{equation}
u_\rrs=-G^{\T\dagger}_\rr u_\rrs G^\T_\rs\label{psg_t}
\end{equation}
exist. Such generalized representations of symmetries are called {\it projective symmetries}. They may differ from the trivial ones if $G^{\mathcal{S}'}_\rr\neq\tau^0$ (due to the gauge transformation already performed, projective time-reversal symmetry differs from trivial time reversal if $G^\T_\rr\neq i\tau^2$). Thus, even if a system seems to violate a symmetry because $u_\rrs\neq u_{\mathcal{S}'(\rr)\mathcal{S}'(\rs)}$, the symmetry may still be intact in the physical spin-1/2 subspace, if there are matrices $G^{\mathcal{S}'}_\rr$ that satisfy Eqs.~(\ref{psg_s}) or (\ref{psg_t}). In other words, Eqs.~(\ref{psg_s}) and (\ref{psg_t}) ensure that the symmetries of the system are fulfilled {\it after} projection. We note that in principle, also the SU(2) spin symmetry needs to be implemented projectively. Such a procedure has been performed in Ref.~\onlinecite{chen12} where it is shown that this generalized symmetry condition allows for a certain class of mean-field Hamiltonians which are not of the form of Eq.~(\ref{mf_hamiltonian_f}). The resulting PSG classification leads to a slightly enhanced number of spin-liquid states. Here, however, we restrict such subtleties and treat the SU(2) spin symmetry non-projectively. In the SU(2) spin symmetry broken case discussed below these additional PSG phases are also included.

We can now formulate the meaning of the IGG in a different way: If $W_\rr$ is a gauge transformation that leaves the mean-field ansatz invariant [i.e., $W_\rr$ fulfills Eq.~(\ref{igg}) and therefore belongs to the IGG] then $W_\rr$ can be interpreted as the gauge transformation of a projective symmetry associated with the {\it identity} transformation. In the $\mathds{Z}_2$ case considered here, Eq.~(\ref{igg}) implies that either $W_\rr\equiv\tau^0$ or $W_\rr\equiv-\tau^0$ on {\it all} sites.

%%%%%%%%%%%%%%%%%%%%%%%%%%%%%%%%%%%%%%%%%%%%%%%%%%%%%%%%%%%%%%
% II.C. CLASSIFICATION OF PSG REPRESENTATIONS
%%%%%%%%%%%%%%%%%%%%%%%%%%%%%%%%%%%%%%%%%%%%%%%%%%%%%%%%%%%%%%

\subsection{Classification of PSG representations}
The PSG approach aims to classify all different {\it gauge inequivalent} mean-field ans\"atze $u_\rrs$ which fulfill the projective symmetries of the system and therefore represent spin-liquid states. In the following, we briefly illustrate how such an analysis is carried out. Instead of directly identifying possible ans\"atze $u_\rrs$, we first classify all allowed sets of gauge transformations $G^\mathcal{S}_\rr$ associated with the lattice transformations and time reversal, $\mathcal{S}=\{\T,T_x,T_y,P_x,P_y,P_{xy}\}$. The actual spin-liquid solutions given by the ans\"atze $u_\rrs$ are then determined by Eqs.~(\ref{psg_s}) and (\ref{psg_t}).

The gauge transformations $G^\mathcal{S}_\rr$ are not independent of each other. Relations among them originate from commutation relations between pairs of symmetry operations. Most pairs $\{\mathcal{S}_a,\mathcal{S}_b\}$ fulfill the simple equality $\mathcal{S}_a^{-1}\mathcal{S}_b^{-1}\mathcal{S}_a\mathcal{S}_b=\mathcal{I}$, where $\mathcal{I}$ is the identity transformation. Exceptions are the following,
\begin{eqnarray}
&&T_x P_x^{-1}T_x P_x=T_y P_y^{-1}T_y P_y=T_x^{-1}P_{xy}^{-1}T_y P_{xy}\notag\\
&=&T_y^{-1}P_{xy}^{-1}T_x P_{xy}=P_x^{-1}P_{xy}^{-1}P_y P_{xy}=P_y^{-1}P_{xy}^{-1}P_x P_{xy}\label{commutators}\notag\\
&=&\mathcal{I}\;.
\end{eqnarray}
Furthermore, the squares of the point-group symmetries and time reversal are given by
\begin{equation}
\T^2=(-1)^{N_\text{f}}\;,\quad P_x^2=P_y^2=P_{xy}^2=\mathcal{I}\;,\label{squares}
\end{equation}
where $N_\text{f}$ is the number of fermions. All these relations consist of two or four successive symmetry operations. For each of these sequences of transformations there is a gauge transformation associated with them. For example, the operation $\mathcal{O}_{ab}=\mathcal{S}_a^{-1}\mathcal{S}_b^{-1}\mathcal{S}_a\mathcal{S}_b=\mathcal{I}$ comes along with the gauge transformation
\begin{equation}
G^{\mathcal{O}_{ab}}_\rr=\left(G^{\mathcal{S}_a}_{\mathcal{S}^{-1}_b \mathcal{S}_a \mathcal{S}_b (\rr)}\right)^\dagger \left(G^{\mathcal{S}_b}_{\mathcal{S}_a \mathcal{S}_b (\rr)}\right)^\dagger G^{\mathcal{S}_a}_{\mathcal{S}_a \mathcal{S}_b (\rr)} G^{\mathcal{S}_b}_{\mathcal{S}_b (\rr)}\;.\label{g_example}
\end{equation}
Since $\mathcal{O}_{ab}$ is the identity transformation, we can now make use of the fact that the IGG is $\mathds{Z}_2$: As noted before, the gauge transformation associated with the identity operation is given by the IGG, i.e., $G^{\mathcal{O}_{ab}}_\rr$ is either $G^{\mathcal{O}_{ab}}_\rr= \tau^0$ or $G^{\mathcal{O}_{ab}}_\rr= -\tau^0$. These two possibilities eventually give rise to two different classes of solutions for the matrices $G^{\mathcal{S}_a}_\rr$ and $G^{\mathcal{S}_b}_\rr$. Similar results are obtained for all relations between the symmetries. In each case the $\mathds{Z}_2$ gauge structure leads to two choices of signs for the net gauge transformation associated with the above sequences of symmetry operations. Different combinations of signs finally define distinct spin-liquid states, yielding a classification scheme on very general grounds.

A closer analysis shows that one can always find a gauge in which the matrices $G^\mathcal{S}_\rr$ have the convenient form
\begin{equation}
G^\mathcal{S}_\rr=d^\mathcal{S}_\rr g_\mathcal{S}\;,\label{spatial_dependence}
\end{equation}
where $d^\mathcal{S}_\rr=\pm1$ is a site-dependent function that defines the spatial variation of $G^\mathcal{S}_\rr$, while $g_\mathcal{S}$ is a (spatially constant) $2\times 2$ SU(2) matrix. Generally, $d^\mathcal{S}_\rr$ follows from the relations in Eq.~(\ref{commutators}) which include translations, while the form of $g_\mathcal{S}$ is determined from the identities between different point-group symmetries and time reversal. Considering the commutations between all symmetry operations, one finds\cite{eta}
\begin{eqnarray}
&G^\T_\rr=\eta_{\T}^{x+y}g_\T\;,\quad G^{T_x}_\rr=\eta^y\tau^0\;,\quad G^{T_y}_\rr=\tau^0\;,&\notag\\
&G^{P_x}_\rr=\eta^x_1\eta^y_2g_{P_x}\;,\quad G^{P_y}_\rr=\eta^x_2\eta^y_1g_{P_y}\;,&\notag\\
&G^{P_{xy}}_\rr=\eta^{xy}g_{P_{xy}}\;,&\label{g_solutions}
\end{eqnarray}
where the exponents on the $\eta$'s contain components of $\rr=(x,y)$. The $\eta$-parameters are given by $\eta_\T=\pm1$, $\eta=\pm1$, $\eta_1=\pm1$, $\eta_2=\pm1$ with uncorrelated signs. Furthermore, the identities among different point-group symmetries and time reversal [Eq.~(\ref{commutators})] and the identities in Eq.~(\ref{squares}) directly relate to equations for $g_\mathcal{S}$,
\begin{eqnarray}
&g^{-1}_\T g^{-1}_{P_x}g_\T g_{P_x}=\pm\tau^0\;,\quad g^{-1}_\T g^{-1}_{P_y}g_\T g_{P_y}=\pm\tau^0\;,&\notag\\
&g^{-1}_\T g^{-1}_{P_{xy}}g_\T g_{P_{xy}}=\pm\tau^0\;,\quad g^{-1}_{P_x}g^{-1}_{P_y}g_{P_x}g_{P_y}=\pm\tau^0\;,&\notag\\
&g^{-1}_{P_x}g^{-1}_{P_{xy}}g_{P_y}g_{P_{xy}}=\pm\tau^0\;,&\notag\\
&g^2_\T=\pm\tau^0,\; g^2_{P_x}=\pm\tau^0,\; g^2_{P_y}=\pm\tau^0,\; g^2_{P_{xy}}=\pm\tau^0.&\label{gggg}
\end{eqnarray}
These equations can be solved, leading to 17 different sets of matrices $g_\T$, $g_{P_x}$, $g_{P_y}$, $g_{P_{xy}}$ shown in Eqs.~(\ref{su2_first}) - (\ref{su2_last}) of Appendix~\ref{g_solutions_1}. Note that not all combinations of signs in Eq.~(\ref{gggg}) yield a solution for the matrices $g_\mathcal{S}$. Taking into account the $2^4=16$ choices for the signs of $\eta_\T$, $\eta$, $\eta_1$, $\eta_2$, there are altogether $16\cdot 17=272$ distinct spin-liquid solutions.\cite{wen02}

We emphasize that the matrices $G^\mathcal{S}_\rr$ are not uniquely defined. In general, a site-dependent gauge shift $u_\rrs\rightarrow W_\rr^\dagger u_\rrs W_\rs$ transforms $G^\mathcal{S}_\rr$ according to $G^\mathcal{S}_\rr\rightarrow W_\rr^\dagger G^\mathcal{S}_\rr W_{\mathcal{S}^{-1}(\rr)}$. The special gauge choice that leads to Eqs.~(\ref{spatial_dependence}) and (\ref{g_solutions}), however, is particularly useful since $G^{T_x}_\rr$ and $G^{T_y}_\rr$ have simple forms. Performing a {\it global} gauge transformation $u_\rrs\rightarrow W^\dagger u_\rrs W$ simultaneously rotates all Pauli matrices in Eqs.~(\ref{su2_first}) - (\ref{su2_last}) but leaves Eq.~(\ref{g_solutions}) unchanged. Furthermore, as a result of the $\mathds{Z}_2$ IGG, the signs of $G^\mathcal{S}_\rr$ are not fixed such that $G^\mathcal{S}_\rr$ may always be changed to $-G^\mathcal{S}_\rr$ for each symmetry $\mathcal{S}$ individually.

%%%%%%%%%%%%%%%%%%%%%%%%%%%%%%%%%%%%%%%%%%%%%%%%%%%%%%%%%%%%%%
% II.D. PSG MEAN-FIELD ANSATZE
%%%%%%%%%%%%%%%%%%%%%%%%%%%%%%%%%%%%%%%%%%%%%%%%%%%%%%%%%%%%%%

\subsection{PSG mean-field ans\"atze}
Having classified all different representations of projective symmetries, we now study the corresponding ans\"atze $u_\rrs$. Evaluating Eq.~(\ref{psg_s}) for $\mathcal{S}=T_x, T_y$ with $G^{T_x}_\rr$ and $G^{T_y}_\rr$ given in Eq.~(\ref{g_solutions}), one finds that $u_\rrs$ may be written as
\begin{equation}
u_\rrs=\eta^{x\dy} u_\drr\;,\label{u_m}
\end{equation}
where we have introduced the matrices $u_\drr$ which only depend on the {\it difference} $\drr\equiv(\dx,\dy)=\rs-\rr$ between sites $\rr$ and $\rs$. This shows that there are two types of solutions: For $\eta=1$ one obtains translation invariant ans\"atze $u_\rrs\equiv u_\drr$. Otherwise, if $\eta=-1$ the ans\"atze $u_\rrs$ form a stripe-like pattern which breaks translation invariance in the $x$-direction (we emphasize again that due to the projective construction, translation invariance is still intact in the physical spin-1/2 subspace). Inserting Eqs.~(\ref{g_solutions}) and (\ref{u_m}) into Eqs.~(\ref{psg_s}) and (\ref{psg_t}) yields further conditions for $u_\drr$,
\begin{eqnarray}
&-\eta_\T^{\dx+\dy}g_\T^\dagger u_\drr g_\T = u_\drr\;,&\notag\\
&\eta^\dx_1\eta^\dy_2 g_{P_x}^\dagger u_{P_x(\drr)} g_{P_x} = u_\drr\;,&\notag\\
&\eta^\dx_2\eta^\dy_1 g_{P_y}^\dagger u_{P_y(\drr)} g_{P_y} = u_\drr\;,&\notag\\
&\eta^{\dx \dy}g_{P_{xy}}^\dagger u_{P_{xy}(\drr)} g_{P_{xy}}=u_\drr\;,&\notag\\
&\eta^{\dx \dy} u_{-\drr}^\dagger =u_\drr\;.&\label{g_conditions}
\end{eqnarray}
Here, the last equation comes from the Hermiticity condition $u_\rrs=u_\rsr^\dagger$. Note that the lattice symmetries act on $\drr$ in the same way as they act on $\rr$ [see Eq.~(\ref{define_symmetries})]. As mentioned before, in the spin-isotropic case $u_\rrs$ has the form of Eq.~(\ref{u_ij2}) or equivalently $u_\drr=i s^0_\drr\tau^0+\sum_{j=1}^3 s^j_\drr \tau^j$. With this decomposition, the conditions in Eq.~(\ref{g_conditions}) can be easily analyzed. As shown in Appendix~\ref{g_solutions_1}, in almost all\cite{almost_all} PSG representations, $g_\mathcal{S}$ is either given by $g_\mathcal{S}=\tau^0$ or by $g_\mathcal{S}=i\tau^j$ ($j=1,2,3$). In these cases, one can use the identity
\begin{equation}
g_\mathcal{S}^\dagger \tau^\mu g_\mathcal{S}=\pm\tau^\mu\quad(\mu=0,1,2,3)\;.
\end{equation}
It follows from Eq.~(\ref{g_conditions}) that $s^\mu_{\mathcal{S}(\drr)}=\pm s^\mu_\drr$, i.e., the action of $\mathcal{S}$ on the coefficient $s^\mu_\drr$ can only change its sign but does not mix components with different $\mu$. Hence, all coefficients $s^\mu_\drr$ with $\dx,\dy\geq 0$ and $\dx\geq \dy$ may be chosen as free parameters, while all other $s^\mu_\drr$ follow from symmetry operations [see Eq.~(\ref{g_conditions})] and only differ by a sign.
%
%%%%%%%%%%%%%%%%%%%%%%%%%%% Figure 1 %%%%%%%%%%%%%%%%%%%%%%%%%%%
%
\begin{figure}[t]
\centering
\includegraphics[width=0.99\linewidth]{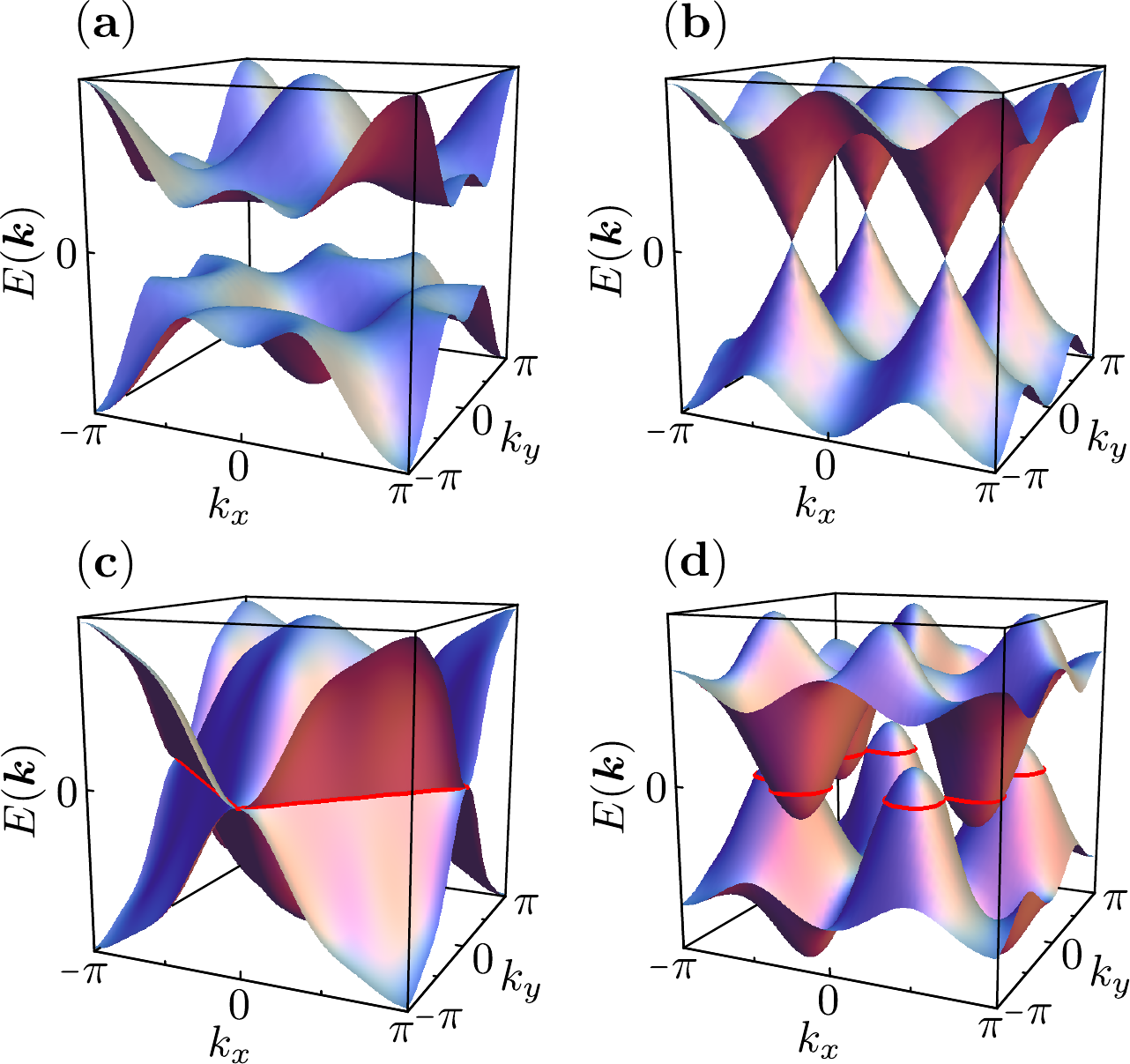}
\caption{Examples for spinon band-structures within different PSG representations when the IGG is $\mathds{Z}_2$. In (a) - (c) the system is spin-isotropic, while in (d) the SU(2) symmetry is broken down to U(1) rotation symmetry around the $z$-axis. (a) Band structure for a fully gapped spin liquid with $\eta=1$, $\eta_\T=\eta_1=\eta_2=-1$ and $g_\T=i\tau^2$, $g_{P_x}=g_{p_y}=i\tau^3$, $g_{P_{xy}}=\tau^0$ [see Eq.~(\ref{case_8})]. In (b) the spinons exhibit a Dirac-cone dispersion with four Dirac points in the first Brillouin zone. This representation is given by $\eta_\T=-1$, $\eta=\eta_1=\eta_2=1$ and $g_\T=g_{P_x}=g_{p_y}=\tau^0$, $g_{P_{xy}}=i\tau^3$ [see Eq.~(\ref{case_3})]. As illustrated in (c) the spinons may also be gapless with a finite Fermi surface (red lines). In this representation we have $\eta_\T=-1$, $\eta=\eta_1=\eta_2=1$ and $g_\T=g_{P_{xy}}=i\tau^2$, $g_{P_x}=g_{p_y}=\tau^0$ [see Eq.~(\ref{case_9})]. The band structure in (d) shows the spinon dispersion in the same representation as in (b) but with the SU(2) spin symmetry broken down to U(1). Such a generalization may qualitatively change the band structure. In the specific example, the Dirac cones become unstable and a finite Fermi surface (red lines) emerges.}
\label{fig:bands_su2_u1}
\end{figure}

Even though we have explicitly enforced $\mathds{Z}_2$ gauge structure in our arguments above, the ans\"atze $u_\rrs$ in certain PSG representations may still ``accidentally" belong to a larger IGG such as U(1) or SU(2). Furthermore, within various PSG representations the mean-field solution $u_\rrs$ even vanishes completely. These cases are artifacts of our simplified mean-field approach. In a generalized treatment which also includes gauge fluctuations and quartic terms in the fermionic Hamiltonian each PSG representation should lead to a finite spin-liquid ansatz with $\mathds{Z}_2$ gauge structure. In total, among the 272 PSG representations we identify 50 instances where $u_\rrs$ vanishes identically. The remaining representations contain 28 cases with SU(2) gauge structure and 4 cases with U(1) gauge structure. Hence, there are altogether $272-50-28-4=190$ finite $\mathds{Z}_2$ PSG mean-field solutions when SU(2) spin symmetry is intact.

Further information about these spin liquids can be obtained from the spinon band structure by transforming $u_\rrs$ into $k$-space. In Ref.~\onlinecite{wen02} various PSG representations have been studied in great detail. Here, we only report the most important properties. In general, the spinon bands of $\mathds{Z}_2$ spin liquids are either fully gapped or gapless. In the latter case, spinons may have a Fermi surface or distinct Fermi points (such as a Dirac-cone dispersion). Examples for band structures in these three cases are shown in Fig.~\ref{fig:bands_su2_u1} (a)-(c). Spin liquids with a gap to all excitations will turn out to be particularly interesting in subsequent sections, as the corresponding spinon band structures may be topological. Since gauge excitations (called visions\cite{wen91}) in $\mathds{Z}_2$ spin liquids are gapped as well, such phases represent rigid quantum states where gauge fluctuations only induce short-range interactions between the spinons. In the following sections we restrict our considerations to $\mathds{Z}_2$ spin liquids where the spinon dispersion is fully gapped (the reasons for this will become clear in Sec.~\ref{no_symmetry}).

%%%%%%%%%%%%%%%%%%%%%%%%%%%%%%%%%%%%%%%%%%%%%%%%%%%%%%%%%%%%%%
% III. U(1) SPIN-ROTATION-INVARIANT Z_2 SPIN LIQUIDS
%%%%%%%%%%%%%%%%%%%%%%%%%%%%%%%%%%%%%%%%%%%%%%%%%%%%%%%%%%%%%%

\section{U(1) spin-rotation-invariant $\mathds{Z}_2$ spin liquids}\label{u1}
We now generalize the PSG method to systems where the SU(2) spin symmetry is broken down to U(1) rotation symmetry around the $z$ axis. While the mean-field Hamiltonian will differ as compared to the previous section, the general procedure remains the same. We first specify naive (i.e. non-projective) implementations of the symmetries. Due to the gauge freedom in our system we then supplement these symmetry operations with additional gauge transformations. Commutation relations between the symmetries pose conditions on the gauge transformations which together with the $\mathds{Z}_2$ IGG lead to a finite number of PSG representations.

In spin space, a U(1) rotation around the $z$ axis is implemented by $f_\rr\rightarrow \exp(-i \sigma^3 \phi/2) f_\rr$  [i.e., $f_{\rr\uparrow}\rightarrow \exp(-i\phi/2)f_{\rr\uparrow}$ and $f_{\rr\downarrow}\rightarrow \exp(i\phi/2)f_{\rr\downarrow}$] where $\phi$ is the polar angle in the $x$-$y$ plane. [For simplicity, we again treat the U(1) spin symmetry non-projectively. As in the SU(2) case, a projective implementation should only lead to minor modifications such as a slightly higher number of PSG representations. We emphasize that in the case of a maximally lifted spin symmetry as studied in the next section, such representations are all covered.] A mean-field ansatz which is invariant under this transformation can only contain terms coupling $f_{\rr\uparrow(\downarrow)}^\dagger$ with $f_{\rr\uparrow(\downarrow)}$ or $f_{\rr\uparrow}^\dagger$ with $f_{\rr\downarrow}^\dagger$. These are exactly the terms of the SU(2) spin symmetric mean-field Hamiltonian in Eq.~(\ref{mf_hamiltonian_full}); however, in the U(1) case there is no constraint that restricts the matrix $u_\rrs$ to the form of Eqs.~(\ref{u_ij}) or (\ref{u_ij2}). Apart from conditions due to the lattice symmetries and time reversal, the new Hamiltonain is therefore given by Eq.~(\ref{mf_hamiltonian_full}), where $u_\rrs$ can be an {\it arbitrary} matrix with $u_\rrs=u_\rsr^\dagger$. We parametrize $u_\rrs$ as $u_\rrs=u^s_\rrs+u^{t_1}_\rrs$ where $u^s_\rrs$ denotes the singlet terms discussed in the last section, $u^s_\rrs=i s^0_\rrs\tau^0+\sum_{j=1}^3 s^j_\rrs \tau^j$; recall Eq.~(\ref{u_ij2}).  The new triplet terms $u_\rrs^{t_1}$ correspond to terms of the form $f^\dagger_{\rr\uparrow}f_{\rs\uparrow}-f^\dagger_{\rr\downarrow}f_{\rs\downarrow}$ (spin-dependent hopping) or $f^\dagger_{\rr\uparrow}f^\dagger_{\rs\downarrow}+f^\dagger_{\rr\downarrow}f^\dagger_{\rs\uparrow}$ (triplet pairing) and may be written as 
\begin{equation}
u^{t_1}_\rrs=t^0_{1,\rrs}\tau^0+i\sum_{j=1}^3 t^j_{1,\rrs} \tau^j\;.\label{ut1_ij}
\end{equation}
Note that all coefficients $s^\mu_\rrs$, $t^{\mu}_{1,\rrs}$ are real.

The set of discrete symmetries that needs to be enforced is the same as in the spin-isotropic case. Particularly, the effect of inversion $P_z$ acting as $z\rightarrow -z$ is still trivial. This can be easily seen: Since the spin operator ${\boldsymbol S}$ is a pseudovector, a three dimensional inversion $\rr\rightarrow -\rr$ leaves ${\boldsymbol S}$ invariant. Hence, the operation $P_z$ in spin space is identical to a $\pi$-rotation in the $x$-$y$ plane with $S_x \rightarrow -S_x$, $S_y \rightarrow -S_y$, $S_z\rightarrow S_z$. As we have constructed our system to be invariant under arbitrary U(1) spin rotations in the $x$-$y$ plane, $P_z$ is redundant and can be omitted in our analysis. There is, however, one important difference as compared to the spin-isotropic case. So far we have only considered the action of the point-group symmetries $P_x$, $P_y$, $P_{xy}$ on the lattice but we have ignored that they also transform the spin. For SU(2) spin symmetric systems this was justified. In the U(1) case discussed here, however, even though the mean-field Hamiltonian is invariant under rotations in the $x$-$y$ plane, its triplet part $u_\rrs^{t_1}$ is {\it not} invariant under (naive) spin reflections $P_x$, $P_y$, $P_{xy}$ in the $x$-$y$ plane ($u_\rrs^{t_1}$ picks up a minus sign). This follows from the fact that the spin ${\boldsymbol S}$ is a pseudovector.

A PSG classification scheme that incorporates the effects of spin transformations is most conveniently carried out in the full four-component Nambu space, spanned by the spinor $\Psi_\rr=(f_{\rr\uparrow},f_{\rr\downarrow}^\dagger,f_{\rr\downarrow},-f_{\rr\uparrow}^\dagger)^\text{T}$. The upper two components of $\Psi_\rr$ are identical to $\psi_\rr$ while the lower two components arise from $\psi_\rr$ under (naive) time reversal. First, in terms of spinors $f_\rr$, the full transformations associated with the point-group symmetries $P_x$, $P_y$, $P_{xy}$ are defined up to a gauge transformation by
\begin{eqnarray}
&P_x:\;f_\rr\rightarrow \exp(-i\pi\sigma^1/2)f_{P_x(\rr)}\;,&\notag\\
&P_y:\;f_\rr\rightarrow \exp(-i\pi\sigma^2/2)f_{P_y(\rr)}\;,&\notag\\
&P_{xy}:\;f_\rr\rightarrow \exp(-i\pi\sigma^1/2)\exp(-i\sigma^3/4)f_{P_{xy}(\rr)}\;.&
\end{eqnarray}
The action on the spinors $\Psi_\rr$ is given by simple $4\times4$ matrices $\mathcal{D_{S''}}$ that couple the upper two and lower two components,
\begin{equation}
\Psi_\rr\rightarrow\mathcal{D_{S''}}\Psi_{\mathcal{S}''(\rr)}\quad\text{with}\quad
\mathcal{D_{S''}}=
\left(\begin{array}{cc}
0 & \gamma_{\mathcal{S}''}\tau^0\\
-\gamma_{\mathcal{S}''}^*\tau^0&0
\end{array}\right)\;,\label{d_s}
\end{equation} 
where the point-group symmetries are denoted by $\mathcal{S}''=\{P_x,P_y,P_{xy}\}$ and $\gamma_{\mathcal{S}''}$ is given by $\gamma_{P_x}=-i$, $\gamma_{P_y}=-1$, $\gamma_{P_{xy}}=(1-i)/\sqrt{2}$. We also rewrite the mean-field Hamiltonian in terms of four-component Nambu spinors $\Psi_\rr$,
\begin{eqnarray}
H_{\text{mf}}&=&\frac{1}{2}\sum_\pair\left( \Psi_\rr^\dagger \tilde{u}_\rrs\Psi_\rs+\text{h.c.}\right)\notag\\
&&+\frac{1}{2}\sum_\rr\sum_{j=1}^3 a_j\Psi_\rr^\dagger
\left(\begin{array}{cc}
\tau^j & 0 \\
0 & \tau^j
\end{array}\right)
\Psi_\rr\label{mf_hamiltonian_new}
\end{eqnarray}
with
\begin{equation}
\tilde{u}_\rrs=
\left(\begin{array}{cc}
u_\rrs^s+u_\rrs^{t_1} & 0 \\
0 & u_\rrs^s-u_\rrs^{t_1} 
\end{array}\right)\;.\label{tilde_u}
\end{equation}
Note that the upper-left and lower-right blocks yield identical contributions [hence the factor $1/2$ in Eq.~(\ref{mf_hamiltonian_new})]. Similarly, gauge transformations in the new basis are performed by
\begin{equation}
\Psi_\rr\rightarrow\tilde{W}\Psi_\rr\quad\text{with}\quad \tilde{W}=
\left(\begin{array}{cc}
W & 0 \\
0 & W
\end{array}\right)\;,\label{gauge_trafo_new}
\end{equation}
where $W$ again denotes a two dimensional SU(2) matrix. Finally, instead of Eq.~(\ref{psg_s}), the defining equation for the PSG representations now also needs to be equipped with a spin transformation $\mathcal{D_S}$ if $\mathcal{S}$ is a point-group symmetry,
\begin{equation}
\tilde{u}_\rrs=\tilde{G}^{\mathcal{S}\dagger}_{\mathcal{S}(\rr)}\mathcal{D}_\mathcal{S}^\dagger \tilde{u}_{\mathcal{S}(\rr)\mathcal{S}(\rs)}\mathcal{D}_\mathcal{S}\tilde{G}^\mathcal{S}_{\mathcal{S}(\rs)}\;;\label{psg_s_new}
\end{equation}
otherwise $\mathcal{D_S}$ is the identity matrix. Here the gauge transformations $\tilde{G}^\mathcal{S}_\rr$ associated with $\mathcal{S}$ have the form of Eq.~(\ref{gauge_trafo_new}).

Having formulated the PSG scheme in the basis of $\Psi_\rr$, we now repeat the classification of spin liquids in the case of U(1) rotation invariance. The first step is again the identification of relations among the symmetries. In our case, all relations in Eqs.~(\ref{commutators}) and (\ref{squares}) remain valid. As in Sec.~\ref{su2}, we consider a sequence of symmetry operations $\mathcal{O}_{ab}=\mathcal{S}_a^{-1}\mathcal{S}_b^{-1}\mathcal{S}_a\mathcal{S}_b=\mathcal{I}$ and determine the corresponding spin and gauge transformations in the extended Nambu space. Incorporating the matrices $\mathcal{D_S}$ (which are unity if $\mathcal{S}$ is not a point-group symmetry) Eq.~(\ref{g_example}) turns into
\begin{eqnarray}
\mathcal{D}_{\mathcal{O}_{ab}}\tilde{G}^{\mathcal{O}_{ab}}_\rr&=&\left(\tilde{G}^{\mathcal{S}_a}_{\mathcal{S}^{-1}_b \mathcal{S}_a \mathcal{S}_b (\rr)}\right)^\dagger \mathcal{D}_{\mathcal{S}_a}^\dagger \left(\tilde{G}^{\mathcal{S}_b}_{\mathcal{S}_a \mathcal{S}_b (\rr)}\right)^\dagger \mathcal{D}_{\mathcal{S}_b}^\dagger\times\notag\\
&&\times  \mathcal{D}_{\mathcal{S}_a} \tilde{G}^{\mathcal{S}_a}_{\mathcal{S}_a \mathcal{S}_b (\rr)}  \mathcal{D}_{\mathcal{S}_b} \tilde{G}^{\mathcal{S}_b}_{\mathcal{S}_b (\rr)}\;.\label{g_example_new}
\end{eqnarray}
Since spin and gauge transformations operate in different subspaces, they commute with each other, leading to
\begin{eqnarray}
\mathcal{D}_{\mathcal{O}_{ab}}\tilde{G}^{\mathcal{O}_{ab}}_\rr&=&\mathcal{D}_{\mathcal{S}_a}^\dagger \mathcal{D}_{\mathcal{S}_b}^\dagger \mathcal{D}_{\mathcal{S}_a} \mathcal{D}_{\mathcal{S}_b}  \left(\tilde{G}^{\mathcal{S}_a}_{\mathcal{S}^{-1}_b \mathcal{S}_a \mathcal{S}_b (\rr)}\right)^\dagger \times\notag\\
&&\times \left(\tilde{G}^{\mathcal{S}_b}_{\mathcal{S}_a \mathcal{S}_b (\rr)}\right)^\dagger\tilde{G}^{\mathcal{S}_a}_{\mathcal{S}_a \mathcal{S}_b (\rr)} \tilde{G}^{\mathcal{S}_b}_{\mathcal{S}_b (\rr)}\;.\label{g_example_new2}
\end{eqnarray}
For all pairs of symmetries $\{\mathcal{S}_a,\mathcal{S}_b\}$, the net spin transformation $\mathcal{D}_{\mathcal{O}_{ab}}=\mathcal{D}_{\mathcal{S}_a}^\dagger \mathcal{D}_{\mathcal{S}_b}^\dagger \mathcal{D}_{\mathcal{S}_a} \mathcal{D}_{\mathcal{S}_b}$ reduces to $\mathcal{D}_{\mathcal{O}_{ab}}=\pm\mathds{1}_{4\times4}$ [this also applies to all other symmetry operations in Eqs.~(\ref{commutators}) and (\ref{squares}) that are not of the form $\mathcal{O}_{ab}=\mathcal{S}_a^{-1}\mathcal{S}_b^{-1}\mathcal{S}_a\mathcal{S}_b$]. Since the gauge structure is $\mathds{Z}_2$, an additional sign due to $\mathcal{D}_{\mathcal{O}_{ab}}=-\mathds{1}_{4\times4}$ is irrelevant such that in total, the spin transformations cancel out entirely. In complete analogy to Sec.~\ref{su2} one obtains $\tilde{G}^{\mathcal{O}_{ab}}_\rr=\pm\mathds{1}_{4\times4}$, leading to identical conditions $G^{\mathcal{O}_{ab}}_\rr=\pm\tau^0$ in the upper-left and lower-right blocks. Hence, all relations in Eqs.~(\ref{g_solutions}) and (\ref{gggg}) as well as their solutions in Appendix~\ref{g_solutions_1} remain valid, resulting in the same 272 PGS representations. A similar observation has also been made for a U(1) spin symmetric PSG classification on the Kagome lattice.\cite{dodds13} Note that Eq.~(\ref{psg_s_new}) does not couple singlet and triplet parts of the mean-field Hamiltonian among each other. Furthermore, since $u_\rrs^s$ is not affected by spin transformations $\mathcal{D_S}$, the conditions in Eq.~(\ref{g_conditions}) still apply, leading to the same singlet solutions $u_\rrs^s$ as in the SU(2) case.

While the classification of spin liquids as well as the mean-field solutions in the singlet sector are not changed by lifting the spin symmetry from SU(2) to U(1), a finite triplet part $u_\rrs^{t_1}$ may modify the properties of an ansatz significantly. In analogy to Eq.~(\ref{u_m}), translation symmetries $T_x$ and $T_y$ again restrict the form of $u_\rrs^{t_1}$,
\begin{equation}
u_\rrs^{t_1}=\eta^{x\dy} u^{t_1}_\drr\label{u_m2}\;.
\end{equation}
Inserting Eqs.~(\ref{g_solutions}), (\ref{d_s}), (\ref{tilde_u}), and (\ref{u_m2}) into Eq.~(\ref{psg_s_new}) yields conditions for $u_\drr^{t_1}$,
\begin{eqnarray}
&-\eta_\T^{\dx+\dy}g_\T^\dagger u_\drr^{t_1} g_\T = u_\drr^{t_1}\;,&\notag\\
&-\eta^\dx_1\eta^\dy_2 g_{P_x}^\dagger u_{P_x(\drr)}^{t_1} g_{P_x} = u_\drr^{t_1}\;,&\notag\\
&-\eta^\dx_2\eta^\dy_1 g_{P_y}^\dagger u_{P_y(\drr)}^{t_1} g_{P_y} = u_\drr^{t_1}\;,&\notag\\
&-\eta^{\dx\dy}g_{P_{xy}}^\dagger u_{P_{xy}(\drr)}^{t_1} g_{P_{xy}}=u_\drr^{t_1}\;,&\notag\\
&\eta^{\dx\dy} \left(u_{-\drr}^{t_1}\right)^\dagger =u_\drr^{t_1}\;.&\label{um_conditions}
\end{eqnarray}
These relations differ from those in Eq.~(\ref{g_conditions}) by additional minus signs in the equations with $\mathcal{S}=P_x,P_y,P_{xy}$. Such signs can be traced back to the fact that any triplet operator is odd under reflections. Given the expansion $u^{t_1}_\drr=t^0_{1,\drr}\tau^0+i\sum_{j=1}^3 t^j_{1,\drr} \tau^j$ [see Eq.~(\ref{ut1_ij})] one can check that as in the spin isotropic case, Eq.~(\ref{um_conditions}) does not relate coefficients $t^\mu_{1,\drr}$ with different $\mu$ among each other [at least if $g_\mathcal{S}=\tau^0$ or $g_\mathcal{S}=i\tau^j$ ($j=1,2,3$), which applies to almost all\cite{almost_all} PSG representations]. Consequently, symmetry operations $\mathcal{S}$ again only affect the sign of the coefficients, i.e., $t^\mu_{1,\mathcal{S}(\drr)}=\pm t^\mu_{1,\drr}$. When compared to the spin-isotropic case, $s^\mu_{\mathcal{S}(\drr)}=\pm s^\mu_\drr$, the sign is reversed if $\mathcal{S}=P_x,P_y,P_{xy}$ such that the singlet solutions $u_\rrs^s$ are qualitatively different from the triplet solutions $u_\rrs^{t_1}$, leading to significant changes in the spinon band-structures. As an example, Fig.~\ref{fig:bands_su2_u1} shows spinon dispersions for the PSG representation with $\eta_\T=-1$, $\eta=\eta_1=\eta_2=1$ and $g_\T=g_{P_x}=g_{p_y}=\tau^0$, $g_{P_{xy}}=i\tau^3$. The Dirac cone dispersion of the singlet part $u_\rrs^s $[see Fig.~\ref{fig:bands_su2_u1} (b)], becomes unstable when triplet terms $u_\rrs^{t_1}$ are added, leading to a finite Fermi surface [see Fig.~\ref{fig:bands_su2_u1} (d)].

In total, among 272 PSG representations we find 48 cases where $u_\rrs^s$ and $u_\rrs^{t_1}$ vanish identically. In all remaining representations, the ans\"atze have $\mathds{Z}_2$ gauge structure, such that there are altogether $272-48=224$ finite $\mathds{Z}_2$ mean-field solutions with U(1) spin-rotation symmetry. 

Given the form of the mean-field Hamiltonian in Eq.~(\ref{mf_hamiltonian_new}) with $\tilde{u}_\rrs=u_\rrs^s\sigma^0+u^{t_1}_\rrs \sigma^3$ (where $\sigma^\mu$ are Pauli matrices in spin space) the spin dependence only comes from triplet terms $\sim\sigma^3$ along the $z$-direction. In other words, in the usual notation\cite{sigrist91,leggett75} for spin-triplet pairing $\psi_\kk i({\boldsymbol d} \cdot{\boldsymbol \sigma})\sigma^2 \psi_{-\kk}$ the $d$-vector always points in the $z$-direction. Consequently, a topological band structure due to spin-orbit coupling is impossible as this would require a winding of the $d$-vector around the full unit sphere. In principle, a non-trivial band topology could still occur in the particle-hole space, i.e., in each block of Eq.~(\ref{tilde_u}) separately. Based on the present analysis we cannot rule out the existence of symmetry protected edge states due to time reversal or lattice symmetries. However, probing the spinon spectra of the PSG ans\"atze for various different mean-field parameters, we could not identify such phases. As we will show in the next section a non-trivial topology in the spinon bands naturally occurs when we completely break the SU(2) spin symmetry.

%%%%%%%%%%%%%%%%%%%%%%%%%%%%%%%%%%%%%%%%%%%%%%%%%%%%%%%%%%%%%%
% IV. Z_2 SPIN LIQUIDS WITHOUT CONTINUOUS SPIN-ROTATION SYMMETRY
%%%%%%%%%%%%%%%%%%%%%%%%%%%%%%%%%%%%%%%%%%%%%%%%%%%%%%%%%%%%%% 

\section{$\mathds{Z}_2$ spin liquids without continuous spin-rotation symmetry}\label{no_symmetry}
We now turn to the most general case where SU(2) spin symmetry is maximally lifted. As we will see, this has drastic consequences on our analysis, yielding a vast number of new PSG representations. To begin with, the mean-field Hamiltonian now consists of {\it all} types of possible quadratic terms -- including spin-flip hopping $f_{\rr\uparrow(\downarrow)}^\dagger f_{\rs\downarrow(\uparrow)}$ and spin-polarized $p$-wave pairing $f_{\rr\uparrow(\downarrow)}^\dagger f_{\rs\uparrow(\downarrow)}^\dagger$. Both terms break U(1) spin-rotation symmetry in the $x$-$y$ plane and were absent in the previous sections. The Hamiltonian again takes the form of Eq.~(\ref{mf_hamiltonian_new}) where the $4\times 4$ matrix $\tilde{u}_\rrs$ containing all mean-field amplitudes now reads as
\begin{equation}
\tilde{u}_\rrs=
\left(\begin{array}{cc}
u_\rrs^s+u_\rrs^{t_1} & u_\rrs^{t_2}+u_\rrs^{t_3} \\
-u_\rrs^{t_2}+u_\rrs^{t_3} & u_\rrs^s-u_\rrs^{t_1} 
\end{array}\right)\;.\label{tilde_u_new}
\end{equation}
Here we have introduced two new triplet sectors $u_\rrs^{t_2}$ and $u_\rrs^{t_3}$ given by 
\begin{eqnarray}
u^{t_2}_\rrs&=&it^0_{2,\rrs}\tau^0+\sum_{j=1}^3 t^j_{2,\rrs} \tau^j\;,\notag\\
u^{t_3}_\rrs&=&t^0_{3,\rrs}\tau^0+i\sum_{j=1}^3 t^j_{3,\rrs} \tau^j\;,
\end{eqnarray}
where all coefficients $t^\mu_{2/3,\rrs}$ are real. Note that the first two rows of $\tilde{u}_\rrs$ represent the most general complex $2\times 4$ matrix with 16 real parameters. The last two rows are completely determined by these entries. This follows from the fact that the upper two and lower two components of $\Psi_\rr$ are related,
\begin{equation}
\Psi_\rr=
\left(\begin{array}{cc}
0 & -i\tau^2\\
i\tau^2 & 0
\end{array}\right)
\left(\Psi_\rr^\text{T}\right)^\dagger\;.
\end{equation}

As in Sec.~\ref{u1} we need to take into account the effect of reflections in spin space, implemented by $\mathcal{D_S}$ [see Eq.~(\ref{d_s})]. Furthermore, gauge transformations in the extended Nambu basis are again given by Eq.~(\ref{gauge_trafo_new}), and the defining equation of the PSG representations has the form of Eq.~(\ref{psg_s_new}). The crucial difference as compared to Secs.~\ref{su2} and \ref{u1} is that inversion $P_z$ may now act non-trivially and needs to be included in our analysis. The action of $P_z$ on the spinors $f_\rr$ is given (up to a gauge transformation) by
\begin{equation}
P_z:\;f_\rr\rightarrow \exp(-i\pi\sigma^3/2)f_\rr\;,
\end{equation} 
or equivalently in terms of $\Psi_\rr$,
\begin{equation}
\Psi_\rr\rightarrow \mathcal{D}_{P_z}\Psi_\rr\quad\text{with}\quad \mathcal{D}_{P_z}=
\left(\begin{array}{cc}
-i\tau^0 & 0\\
0&i\tau^0
\end{array}\right)\;.\label{pz}
\end{equation}
Hence, when applied to the mean-field Hamiltonian $P_z$ effectively transforms the ansatz $\tilde{u}_\rrs$ [see Eq.~(\ref{tilde_u_new})] via
\begin{equation}
\tilde{u}_\rrs\rightarrow\mathcal{D}^\dagger_{P_z} \tilde{u}_\rrs \mathcal{D}_{P_z}=
\left(\begin{array}{cc}
u_\rrs^s+u_\rrs^{t_1} & -u_\rrs^{t_2}-u_\rrs^{t_3} \\
u_\rrs^{t_2}-u_\rrs^{t_3} & u_\rrs^s-u_\rrs^{t_1} 
\end{array}\right)\;,\label{u_transform_pz}
\end{equation}
where the signs are flipped in the off-diagonal $2\times 2$ blocks. This has interesting consequences: When the off-diagonal blocks are finite, invariance under $z\rightarrow-z$ can only be intact if $P_z$ is implemented {\it projectively},
\begin{equation}
\tilde{u}_\rrs=\tilde{G}^{P_z\dagger}_\rr\mathcal{D}_{P_z}^\dagger \tilde{u}_\rrs\mathcal{D}_{P_z}\tilde{G}^{P_z}_\rs\;,\label{psg_pz}
\end{equation}
with a non-trivial gauge transformation $G^{P_z}_\rr\neq\tau^0$. In other words, there are two types of PSG representations when SU(2) spin symmetry is maximally lifted. First, a representation may be given by $G^{P_z}_\rr\equiv\tau^0$. In this case, the triplet sectors $u_\rrs^{t_2}$ and $u_\rrs^{t_3}$ vanish and $P_z$ acts trivially -- precisely as in the SU(2) and U(1) spin-symmetric cases. This immediately leads to the 272 PSG representations already discussed in the previous sections. These solutions may therefore also be interpreted as PSG representations of a system without spin-rotation symmetries, but that nevertheless retain an {\it accidental} U(1) symmetry on the mean-field level. We will not discuss such solutions again, instead focusing on the far more interesting second case where $G^{P_z}_\rr$ is non-trivial, i.e., $G^{P_z}_\rr\neq\tau^0$. Finite off-diagonal blocks $\pm u_\rrs^{t_2}+u_\rrs^{t_3}$ are then allowed by symmetries, leading to additional PSG representations with novel types of ans\"atze.

In the following, we briefly outline how these new representations are obtained. The extended set of symmetries that needs to be enforced -- now also including $P_z$ -- leads to new commutation relations in addition to Eq.~(\ref{commutators}). These new relations have the form $\mathcal{S}^{-1}P_z^{-1}\mathcal{S}P_z=\mathcal{I}$ where $\mathcal{S}$ can be any of the symmetries $\mathcal{S}=\{\T,T_x,T_y,P_x,P_y,P_{xy}\}$. Furthermore, we have $P_z^2=\mathcal{I}$. Following the arguments after Eq.~(\ref{g_example}), these sequences of symmetries are associated with a net gauge transformation that must be an element of the IGG. As explained below Eq.~(\ref{g_example_new}) gauge transformations $G^\mathcal{S}_\rr$ and spin transformations $\mathcal{D_S}$ commute such that one obtains the relations
\begin{equation}
G^{\mathcal{S}\dagger}_{\mathcal{S}(\rr)}G^{P_z\dagger}_{\mathcal{S}(\rr)}G^\mathcal{S}_{\mathcal{S}(\rr)}G^{P_z}_\rr=\pm\tau^0\quad,\quad\left(G^{P_z}_\rr\right)^2=\pm\tau^0\;.\label{gpz_relations}
\end{equation}
As in Eq.~(\ref{spatial_dependence}), $G^{P_z}_\rr$ may be written in the form $G^{P_z}_\rr=d^{P_z}_\rr g_{P_z}$, where the spatial dependence of $d^{P_z}_\rr$ is determined by Eq.~(\ref{gpz_relations}) if $\mathcal{S}$ is a translation symmetry. It follows that
\begin{equation}
G^{P_z}_\rr=\eta_z^{x+y}g_{P_z}\;,\label{gpz_spatial_dependence}
\end{equation}
where $\eta_z=\pm1$. Inserting Eq.~(\ref{gpz_spatial_dependence}) into Eq.~(\ref{gpz_relations}) and using Eq.~(\ref{g_solutions}) yields the following conditions for $g_{P_z}$ when $\mathcal{S}$ is time-reversal or a point-group symmetry,
\begin{eqnarray}
&g_\T^{-1}g_{P_z}^{-1} g_\T g_{P_z}=\pm\tau^0\;,\quad g_{P_x}^{-1}g_{P_z}^{-1} g_{P_x} g_{P_z}=\pm\tau^0\;,&\notag\\
&g_{P_y}^{-1}g_{P_z}^{-1} g_{P_y} g_{P_z}=\pm\tau^0\;,\quad g_{P_{xy}}^{-1}g_{P_z}^{-1} g_{P_{xy}} g_{P_z}=\pm\tau^0\;,&\notag\\
&g_{P_z}^2=\pm\tau^0\;.&\label{gggg2}
\end{eqnarray}
Here, the different signs are uncorrelated, but not all combinations of signs yield a solution for the matrices $g_\mathcal{S}$. Together with Eq.~(\ref{gggg}), these relations determine all possible PSG representations. As shown in Appendix~\ref{g_solutions_2} there are 55 different sets of matrices $\{g_\T,g_{P_x},g_{P_y},g_{P_{xy}},g_{P_z}\}$ that fulfill Eqs.~(\ref{gggg}) and (\ref{gggg2}). Taking into account the $2^5=32$ combinations of different signs for $\eta_\T$, $\eta$, $\eta_1$, $\eta_2$, and $\eta_z$ there are altogether $32\cdot55=1760$ PSG representations. As mentioned above, 272 of them have $G^{P_z}_\rr\equiv\tau^0$ (i.e., $\eta_z=1$ {\it and} $g_{P_z}=\tau^0$) resulting in an accidental U(1) rotation symmetry. Hence, we find $1760-272=1488$ new PSG representations with a non-trivial action of $P_z$.

Before studying some of these 1488 solutions in more detail (see next section), we first point out several general properties of the ans\"atze $\tilde{u}_\rrs$. The U(1) spin symmetric sectors $u_\rrs^s$ and $u_\rrs^{t_1}$ are again determined by Eqs.~(\ref{g_conditions}) and (\ref{um_conditions}). However, there are now additional conditions resulting from the invariance under $P_z$. Inserting Eqs.~(\ref{tilde_u_new}), (\ref{pz}), and (\ref{gpz_spatial_dependence}) into Eq.~(\ref{psg_pz}) yields
\begin{equation}
\eta_z^{\dx+\dy}g_{P_z}^\dagger u_\drr^s g_{P_z}=u_\drr^s\;,\quad\eta_z^{\dx+\dy}g_{P_z}^\dagger u_\drr^{t_1} g_{P_z}=u_\drr^{t_1}\;.
\end{equation}
These conditions may lead to vanishing coefficients $s_\rrs^\mu$, $t_{1,\rrs}^\mu$ which would otherwise be finite. Most importantly, however, the new ans\"atze may contain finite triplet terms $u_\rrs^{t_2}$ and $u_\rrs^{t_3}$. In analogy to Eqs.~(\ref{u_m}) and (\ref{u_m2}) the form of $G^{T_x}_\rr$ and $G^{T_y}_\rr$ in Eq.~(\ref{g_solutions}) determines the spatial dependence of $u_\rrs^{t_2}$ and $u_\rrs^{t_3}$,
\begin{equation}
u_\rrs^{t_2}=\eta^{x\dy} u^{t_2}_\drr\;,\quad u_\rrs^{t_3}=\eta^{x\dy} u^{t_3}_\drr\label{u_m3}\;.
\end{equation}
Conditions for $u^{t_2}_\drr$ and $u^{t_3}_\drr$ are obtained by inserting the general mean-field ansatz $\tilde{u}_\rrs$ from Eqs.~(\ref{tilde_u_new}) and (\ref{u_m3}) as well as the spin transformations $\mathcal{D_S}$ [Eqs. (\ref{d_s}), (\ref{pz})] into Eq.~(\ref{psg_s_new}). Further using Eqs.~(\ref{g_solutions}) and (\ref{gpz_spatial_dependence}) leads to
\begin{eqnarray}
&-\eta_\T^{\dx+\dy}g_\T^\dagger u_\drr^{t_2} g_\T = u_\drr^{t_2}\;,&\notag\\
&-\eta^{\dx}_1\eta^{\dy}_2 g_{P_x}^\dagger u_{P_x(\drr)}^{t_2} g_{P_x} = u_\drr^{t_2}\;,&\notag\\
&\eta^{\dx}_2\eta^{\dy}_1 g_{P_y}^\dagger u_{P_y(\drr)}^{t_2} g_{P_y} = u_\drr^{t_2}\;,&\notag\\
&-i\eta^{\dx \dy}g_{P_{xy}}^\dagger u_{P_{xy}(\drr)}^{t_2} g_{P_{xy}}=u_\drr^{t_3}\;,&\notag\\
&-\eta_z^{\dx+\dy}g_{P_z}^\dagger u_\drr^{t_2} g_{P_z}=u_\drr^{t_2}\;,&\notag\\
&-\eta^{\dx \dy} \left(u_{-\drr}^{t_2}\right)^\dagger =u_\drr^{t_2}\;,&\label{um2_conditions}
\end{eqnarray}
and
\begin{eqnarray}
&-\eta_\T^{\dx+\dy}g_\T^\dagger u_\drr^{t_3} g_\T = u_\drr^{t_3}\;,&\notag\\
&\eta^{\dx}_1\eta^{\dy}_2 g_{P_x}^\dagger u_{P_x(\drr)}^{t_3} g_{P_x} = u_\drr^{t_3}\;,&\notag\\
&-\eta^{\dx}_2\eta^{\dy}_1 g_{P_y}^\dagger u_{P_y(\drr)}^{t_3} g_{P_y} = u_\drr^{t_3}\;,&\notag\\
&i\eta^{\dx \dy}g_{P_{xy}}^\dagger u_{P_{xy}(\drr)}^{t_3} g_{P_{xy}}=u_\drr^{t_2}\;,&\notag\\
&-\eta_z^{\dx+\dy}g_{P_z}^\dagger u_\drr^{t_3} g_{P_z}=u_\drr^{t_3}\;,&\notag\\
&\eta^{\dx \dy} \left(u_{-\drr}^{t_3}\right)^\dagger =u_\drr^{t_3}\;.&\label{um3_conditions}
\end{eqnarray}
In analogy to the sectors $u_\drr^s$ and $u_\drr^{t_1}$, when expanding $u^{t_2}_\drr=it^0_{2,\drr}\tau^0+\sum_{j=1}^3 t^j_{2,\drr} \tau^j$ and $u^{t_3}_\drr=t^0_{3,\drr}\tau^0+i\sum_{j=1}^3 t^j_{3,\drr} \tau^j$, these relations do not couple coefficients $t_{2/3,\drr}^\mu$ with different $\mu$ among each other [at least if $g_\mathcal{S}=\tau^0$ or $g_\mathcal{S}=i\tau^j$ ($j=1,2,3$), which again applies to nearly all\cite{almost_all} PSG representations].

There is, however, an important difference as compared to the SU(2) and U(1) spin symmetric cases. As shown in Eqs.~(\ref{um2_conditions}) and (\ref{um3_conditions}), the symmetry $P_{xy}$ connects the $t_2$ and the $t_3$ triplet sectors, yielding
\begin{equation}
t^\mu_{3,\drr}=\pm t^\mu_{2,P_{xy}(\drr)}\;.\label{coupling_t2_t3}
\end{equation}
This relation has interesting consequences which we now discuss in more detail. For simplicity let us only consider components with $\mu=1$ and also neglect $u_\drr^s$ and $u_\drr^{t_1}$. Assuming that for a given $\drr$ the ansatz $\tilde{u}_\drr$ has a finite $u_\drr^{t_3}$ part in the $\mu=1$ sector, i.e.,
\begin{equation}
\tilde{u}_\drr=t^1_{3,\drr}
\left(\begin{array}{cc}
0 & i\tau^1\\
i\tau^1 & 0
\end{array}\right)\;,\label{um_t3}
\end{equation} 
it follows from Eq.~(\ref{coupling_t2_t3}) that the $u_{P_{xy}(\drr)}^{t_2}$ part of $\tilde{u}_{P_{xy}(\drr)}$ has -- up to a sign -- the same coefficient,
\begin{equation}
\tilde{u}_{P_{xy}(\drr)}=\pm t^1_{3,\drr}
\left(\begin{array}{cc}
0 & \tau^1\\
-\tau^1 & 0
\end{array}\right)\;.\label{um_t2}
\end{equation} 
Expressing the corresponding mean-field Hamiltonian in terms of $f_{\rr\alpha}$ operators and summing over different $\drr$ (i.e., different pairs of sites) leads to
\begin{eqnarray}
H_\text{mf}\negthickspace&=&\negthickspace\sum_\pair\left[it^1_{3,\drr}\left(-f_{\rr\uparrow}^\dagger f_{\rs\uparrow}^\dagger+f_{\rr\downarrow}^\dagger f_{\rs\downarrow}^\dagger+\text{h.c.}\right)\right.\notag\\
&\pm& t^1_{3,\drr}\negthickspace\left.\left(f_{P_{xy}(\rr)\uparrow}^\dagger f_{P_{xy}(\rs)\uparrow}^\dagger+f_{P_{xy}(\rr)\downarrow}^\dagger f_{P_{xy}(\rs)\downarrow}^\dagger+\text{h.c.}\right)\right]\notag\\
&+&\ldots\;,
\end{eqnarray}
where we have assumed $\drr=\rs-\rr$. Note that the first line contains the term in Eq.~(\ref{um_t3}) while the second line corresponds to Eq.~(\ref{um_t2}). Transforming $H_\text{mf}$ into $k$-space using $f_{\rr\alpha}=\frac{1}{\sqrt{N}}\sum_\kk e^{i\kk\rr}f_{\kk\alpha}$ ($N$ is the total number of lattice sites) yields
\begin{eqnarray}
H_\text{mf}&=&-\sum_\kk t^1_{3,\drr}[\sin(k_x)\pm i\sin(k_y)]\left(f_{\kk\uparrow}^\dagger f_{-\kk\uparrow}^\dagger+\text{h.c.}\right)\notag\\
&&-\sum_\kk t^1_{3,\drr}[\sin(k_x)\mp i\sin(k_y)]\left(f_{-\kk\downarrow}^\dagger f_{\kk\downarrow}^\dagger+\text{h.c.}\right)\notag\\
&&+\cdots\;.
\end{eqnarray}
Here, we only show the terms for $\drr=(1,0)$, while the other contributions are indicated by the ellipsis. As can be seen, the interplay between the $t_2$ and $t_3$ triplet sectors directly generates $p_x\pm ip_y$ pairing with a chirality determined by the sign in Eq.~(\ref{coupling_t2_t3}). Furthermore, the two spin orientations have opposite chirality. This type of time-reversal-invariant `superconductivity'\cite{qi09,qi10,schnyder08} possibly leads to a non-trivial topology in the spinon bands associated with gapless edge states. The properties of such edge states within various PSG representations will be discussed in more detail below. 

It is worth emphasizing once more the role of different triplet pairing terms within our PSG analysis. While triplet pairing of the form $f^\dagger_{\rr\uparrow}f^\dagger_{\rs\downarrow}+f^\dagger_{\rr\downarrow}f^\dagger_{\rs\uparrow}$ (which corresponds to the $d_z$-component in $d$-vector notation) is allowed in the U(1) spin symmetric case, spin-polarized terms such as $f_{\rr\uparrow}^\dagger f_{\rs\uparrow}^\dagger$ and $f_{\rr\downarrow}^\dagger f_{\rs\downarrow}^\dagger$ lift the U(1) spin-rotation symmetry in the $x$-$y$ plane. (Note that in the $d$-vector notation $f_{\rr\uparrow}^\dagger f_{\rs\uparrow}^\dagger$ and $f_{\rr\downarrow}^\dagger f_{\rs\downarrow}^\dagger$ correspond to the combinations $-d_x-id_y$ and $d_x-id_y$, respectively.) As shown above, these terms occur in the U(1) broken case and automatically come along with a pairing amplitude $\sim p_x\pm ip_y$. This behavior of the gap function is specific to systems without spin-rotation symmetries and does not appear when SU(2) or U(1) symmetries are intact.

We finally mention that not all 1488 PSG representations with a non-trivial action of $P_z$ have a finite ansatz $\tilde{u}_{\rrs}$ with $\mathds{Z}_2$ gauge structure. As an artifact of the mean-field treatment, we find 226 cases where $\tilde{u}_\rrs$ vanishes identically. Among the remaining representations there are 28 instances with SU(2) gauge structure and 56 instances with a U(1) gauge structure. Hence, we identify $1488-226-28-56=1178$ finite mean-field ans\"atze with $\mathds{Z}_2$ gauge structure where SU(2) spin symmetry is maximally lifted. 
%
%%%%%%%%%%%%%%%%%%%%%%%%%%% Figure 2 %%%%%%%%%%%%%%%%%%%%%%%%%%%
%
\begin{figure}[t]
\centering
\includegraphics[width=0.99\linewidth]{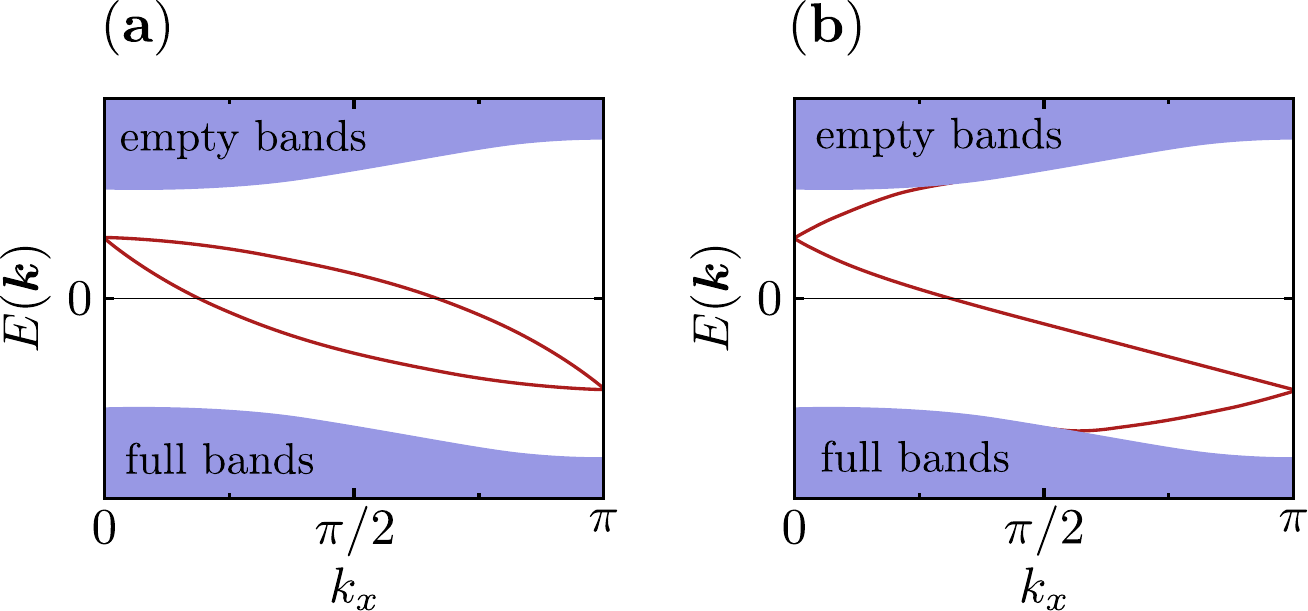}
\caption{Two topologically distinct ways of connecting boundary modes between the time-reversal invariant momenta $k_x=0$ and $k_x=\pi$, following Refs.~\onlinecite{fu07,fu07_2,hasan10,qi11}. In (a) the edge states are connected pairwise, leading to a band structure which is equivalent to a trivial insulator. As shown in (b), the modes at $k_x=0$ and $k_x=\pi$ may also be connected such that there is an odd number of Fermi points. Time reversal preserving deformations of the bands cannot generate a gap in the edge-state spectrum. These two cases define a $\mathds{Z}_2$ topological index for time-reversal invariant band structures in two dimensions. Note that (a) and (b) only illustrate half of the first Brillouin zone. The other half between $k_x=-\pi$ and $k_x=0$ it given by the mirror image.}
\label{fig:z2_invariant}
\end{figure}
%

%%%%%%%%%%%%%%%%%%%%%%%%%%%%%%%%%%%%%%%%%%%%%%%%%%%%%%%%%%%%%%
% V. EXAMPLES FOR Z_2 SPIN LIQUIDS WITHOUT CONTINUOUS SPIN-ROTATION SYMMETRY
%%%%%%%%%%%%%%%%%%%%%%%%%%%%%%%%%%%%%%%%%%%%%%%%%%%%%%%%%%%%%%

\section{Examples for $\mathds{Z}_2$ spin liquids without continuous spin-rotation symmetry}\label{examples}
In the last section we pointed out that a PSG analysis for systems without spin-rotation symmetries naturally captures helical $p_x\pm ip_y$ pairing for the spinons. This property is particularly interesting in the case of gapped $\mathds{Z}_2$ spin liquids as it might lead to topological spinon band structures. To gain more insight into the topology of the spinon bands, we now study various PSG representations with a bulk gap in the spinon excitations and especially focus on possible edge states. It will turn out that for many PSG representations one may define simple $\mathds{Z}_2$ topological invariants for the spinon bands, based on time-reversal symmetry. 

Following the arguments of Refs.~\onlinecite{fu07,fu07_2,hasan10,qi11}, we first briefly recapitulate the role of time-reversal symmetry for the characterization of topological band structures in {\it ordinary} fermionic systems (in which the symmetries are not implemented projectively). In such systems time reversal $\T$ is an antiunitary operator (i.e., $\T c=c^*\T$ for constant $c$) with $\T^2|\phi\rangle=-|\phi\rangle$ when acting on single-particle states $|\phi\rangle$. Given these two properties, Kramer's theorem states that all single-particle eigenstates must at least be two-fold degenerate. One can show that based on this degeneracy all two dimensional time-reversal invariant band structures with a bulk gap can be characterized by a $\mathds{Z}_2$ topological invariant. This may be illustrated by considering a lattice system with cylinder geometry, where $x$ is the cyclic coordinate along the cylinder edge. The first Brillouin zone $k_x\in[-\pi,\pi]$ contains two momenta $k_x=0$ and $k_x=\pi$ that remain invariant under time-reversal. According to Kramers theorem, the spectrum consists of two-fold degenerate pairs of states at these points. For possible boundary modes residing in the bulk gap there are consequently two ways of connecting the pairs of edge states between $k_x=0$ and $k_x=\pi$. As shown in Fig.~\ref{fig:z2_invariant} the Kramer's pairs may be either connected pairwise or in a way that there is an {\it odd} number of crossing points at each fixed energy. While the first case is equivalent to a trivial insulator, the second case represents a topological band structure with time-reversal-protected boundary modes. The number of crossing points modulo 2 for $k_x\in[0,\pi]$ therefore defines a $\mathds{Z}_2$ topological invariant for bulk-gapped, time-reversal-invariant band structures in two dimensions. In Fig.~\ref{fig:z2_invariant}(a) and (b) this index is given by $\nu=0$ and $\nu=1$, respectively. This classification holds for topological insulators\cite{kane05,bernevig06,hasan10} as well as for time-reversal-invariant topological superconductors.\cite{qi09,qi10,schnyder08} In the latter case, however, the edge states are Majorana fermions, reflecting the fact that positive and negative energies are related.

There are various ways of determining $\nu$ for a given band structure. If the spin up and spin down sectors of the Hamiltonian decouple, the calculation of $\nu$ is particularly simple. Denoting the Chern numbers of the two sectors by $n_\uparrow$ and $n_\downarrow$, respectively, $\nu$ can be obtained via\cite{sheng06,hasan10,roy09,konig07}
\begin{equation}
\nu=\frac{n_\uparrow-n_\downarrow}{2}\;\text{mod}\;2\;.\label{z2_invariant}
\end{equation}
We will often use this identity below. It is important to emphasize that the existence of a $\mathds{Z}_2$ topological invariant is a consequence of the fact that $\T$ is an antiunitary operator with $\T^2|\phi\rangle=-|\phi\rangle$. Since no other lattice symmetry has these two properties, time reversal plays a special role among all symmetries of a system.

Within the PSG approach, symmetries are implemented projectively. Such generalized symmetries may be different from the ones of an ordinary fermionic system. Specifically, if $\eta_\T=1$ and $g_\T=i\tau^2$ projective time-reversal symmetry is identical to ordinary time-reversal symmetry, leading to a $\mathds{Z}_2$ topological invariant for two dimensional band structures, as discussed above. Note that even if $g_\T$ is given by an arbitrary linear combination $g_\T=ix_1\tau^1+ix_2\tau^2+ix_3\tau^3\neq i\tau^2$, one can always perform a global gauge rotation $g_\T\rightarrow W^\dagger g_\T W$ such that $g_\T=i\tau^2$. To simplify the discussion of time-reversal symmetry, the PSG solutions listed in Appendix~\ref{g_solutions_1} are all in a gauge where either $g_\T=\tau^0$ or $g_\T=i\tau^2$. While the $\mathds{Z}_2$ classification of time-reversal-invariant band structures remains valid when $\eta_\T=-1$ and $g_\T=i\tau^2$ (see Secs.~\ref{sl696} and~\ref{sl101}), there is no Kramer's theorem for a PSG representation with $g_\T=\tau^0$ and consequently no $\mathds{Z}_2$ invariant based on time-reversal symmetry (see Sec.~\ref{sl504}).

Given this line of arguments, the topological protection of edge states in all the examples presented below can ultimately be traced back to time-reversal invariance. However, we will see that the spatial symmetries can be important as well. In particular, if the projective representations of spatial symmetries are different compared to ordinary fermionic systems they may lead to an {\it additional} protection of edge states. In these cases, the boundary modes can only be gapped out if time-reversal symmetry and certain lattice symmetries are violated {\it simultaneously}. In Secs.~\ref{sl81}, \ref{sl101}, and \ref{sl628} we study PSG  representations where the system effectively decouples into an {\it even} number of sectors in the mean-field Hamiltonian, each one characterized by a non-trivial $\mathds{Z}_2$ invariant based on time-reversal symmetry. Using the arguments above this would imply a trivial $\mathds{Z}_2$ index for the full system. However, we will show that as an effect of spatial symmetries the topological $\mathds{Z}_2$ index of an individual sector may survive, leading to an overall protection that relies on various different symmetries. Note that in this work we do not consider $\mathds{Z}_2$ invariants directly arising from spatial symmetries such as the ones discussed in the context of topological crystalline insulators; see e.g., Ref.~\onlinecite{fu11,teo13,alexandradinata14}. 

In the following subsections we study the spinon band structures in various representative PSG spin-liquid solutions. Choosing specific sets of mean-field parameters, the excitation spectra are calculated for a system on a torus and on a cylinder (if not stated otherwise the cylinder edges are along the $x$-direction). In the latter case, we investigate the origin of the gaplessness of possible boundary modes and define the corresponding topological invariants. We emphasize that the edge-state spectra presented below are not particular to these specific examples but may occur in various other PSG representations in a qualitatively similar way (for example, the edge-state structure shown in Fig.~\ref{fig:psg81} also appears in the PSG representations of Secs.~\ref{sl625} and \ref{sl101}). Furthermore, depending on the precise choice of mean-field parameters a single PSG solution may host various topologically distinct phases (for instance, the PSG representation studied in Sec.~\ref{sl81} also exhibits a parameter regime without edge states). To facilitate the analysis, we study very simple mean-field ans\"atze avoiding longer-ranged hopping and pairing amplitudes. Such ans\"atze may have a gauge group larger than $\mathds{Z}_2$. In each case, however, one may add further terms to the Hamitonian which break the gauge group down to $\mathds{Z}_2$ but do not affect the system's topological properties.

%%%%%%%%%%%%%%%%%%%%%%%%%%%%%%%%%%%%%%%%%%%%%%%%%%%%%%%%%%%%%%
% V.A. EXAMPLE 1
%%%%%%%%%%%%%%%%%%%%%%%%%%%%%%%%%%%%%%%%%%%%%%%%%%%%%%%%%%%%%%

\subsection{PSG with $\{\eta_\T,\eta,\eta_1,\eta_2,\eta_z\}=\{1,1,1,1,1\}$ and $\{g_\T,g_{P_x},g_{P_y},g_{P_{xy}},g_{P_z}\}=\{i\tau^2,\tau^0,\tau^0,\tau^0,i\tau^3\}$}\label{sl625}
In the simplest possible case of a PSG representation without any continuous spin-rotation invariance, the projective symmetries $\T$, $T_x$, $T_y$, $P_x$, $P_y$, and $P_{xy}$ are all identical to their counterparts in ordinary fermion systems. This requires that $\eta=\eta_1=\eta_2=1$ and $g_{P_x}=g_{P_y}=g_{P_{xy}}=\tau^0$. Furthermore, time-reversal symmetry must be implemented by $\eta_\T=1$ and $g_\T=i\tau_2$. As explained above, one can then define a $\mathds{Z}_2$ topological invariant for the spinon band structure. Note that inversion $P_z$ plays a special role: According to our assumption that the ansatz lifts spin-rotation symmetries, $P_z$ must be implemented nontrivially, either by $\eta_z=-1$ or by $g_{P_z}\neq\tau^0$. In this section we study a PSG representation with $\eta_z=1$ and $g_{P_z}=i\tau^3$. Given an ansatz of the form of Eq.~(\ref{tilde_u_new}) with
\begin{eqnarray}
&u^s_\drr=i s^0_\drr\tau^0+\sum_{j=1}^3 s^j_\drr \tau^j\;,&\notag\\
&u^{t_1}_\drr=t^0_{1,\drr}\tau^0+i\sum_{j=1}^3 t^j_{1,\drr} \tau^j\;,&\notag\\
&u^{t_2}_\drr=it^0_{2,\drr}\tau^0+\sum_{j=1}^3 t^j_{2,\drr} \tau^j\;,&\notag\\
&u^{t_3}_\drr=t^0_{3,\drr}\tau^0+i\sum_{j=1}^3 t^j_{3,\drr} \tau^j\;,&
\end{eqnarray}
we choose the mean-field parameters as
\begin{eqnarray}
&s_{\drr=(1,0)}^3=s_{\drr=(0,1)}^3=1\;,&\notag\\
&t_{2,\drr=(1,0)}^1=-t_{3,\drr=(0,1)}^1=1\;,&\notag\\
&s_{\drr=(1,1)}^3=s_{\drr=(1,-1)}^3=0.5\;;&\label{sl625_parameters}
\end{eqnarray}
all other parameters vanish. In the PSG representation discussed here, the spin up and spin down sectors decouple within the present gauge convention. Hence, it will be convenient to use a new basis $\hat{\Psi}_\rr=(f_{\rr\uparrow},f_{\rr\uparrow}^\dagger,f_{\rr\downarrow},f_{\rr\downarrow}^\dagger)^\text{T}$ which groups together $\uparrow$- and $\downarrow$-operators, resulting in a bock-diagonal Hamiltonian. Using the parameters in Eq.~(\ref{sl625_parameters}) and transforming the ansatz into $k$-space yields the Hamiltonian
\begin{equation}
H_\text{mf}=\frac{1}{2}\sum_\kk\hat{\Psi}_\kk^\dagger
\left[\left(\begin{array}{cc}
h_\kk & 0 \\
0 & h^*_{-\kk}
\end{array}\right)+\sum_j^3 a_j
\left(\begin{array}{cc}
\tau^j & 0 \\
0 & \tau^j
\end{array}\right)\right]
\hat{\Psi}_\kk\label{sl625_ham1}
\end{equation}
with
\begin{widetext}
\begin{equation}
h_\kk=
\left(\begin{array}{cc}
\cos(k_x)+\cos(k_y)+\cos(k_x)\cos(k_y) & -i\sin(k_x)-\sin(k_y)\\
i\sin(k_x)-\sin(k_y) & -\cos(k_x)-\cos(k_y)-\cos(k_x)\cos(k_y)
\end{array}\right)\;.\label{sl625_ham2}
\end{equation}
\end{widetext}
Here $a_j$ are Lagrange multipliers enforcing the single occupancy constraint on average. Generally, Lagrange multipliers can be determined from the condition\cite{wen02} $\partial E_\text{g}/\partial a_j=0$, where $E_\text{g}$ is the ground-state energy. Solving this equation numerically for the parameters in Eq.~(\ref{sl625_parameters}), we find\cite{lagrange} $a_1=a_2=0$ and $a_3\approx -0.607$.
%
%%%%%%%%%%%%%%%%%%%%%%%%%%% Figure 3 %%%%%%%%%%%%%%%%%%%%%%%%%%%
%
\begin{figure}[t]
\centering
\includegraphics[width=0.99\linewidth]{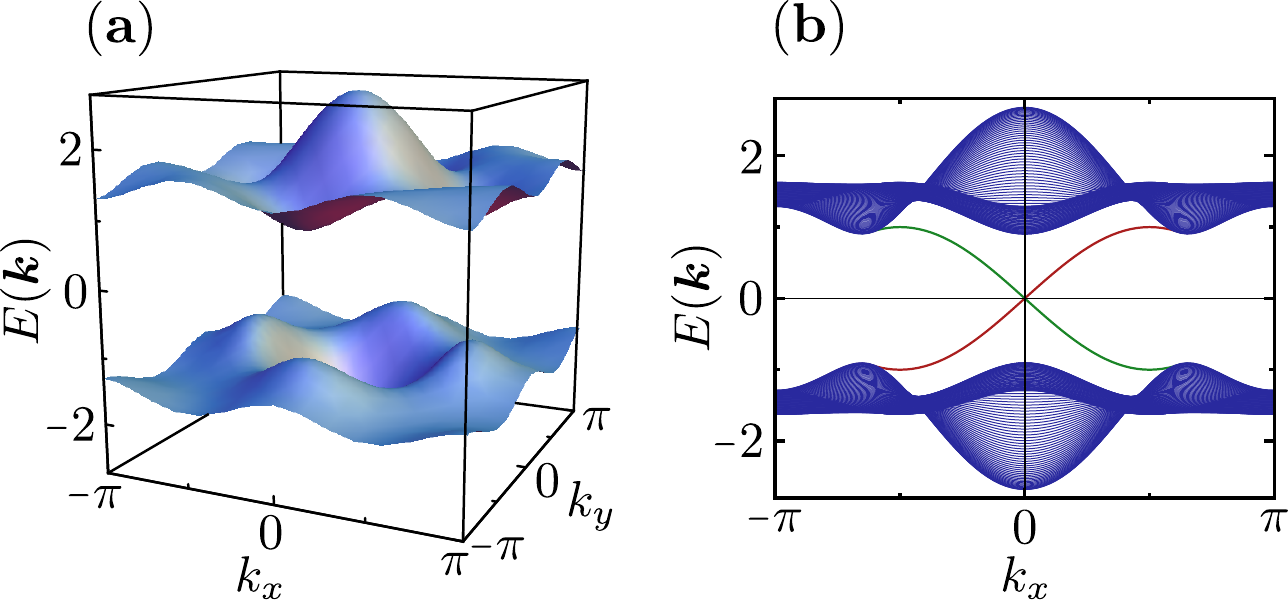}
\caption{Band structure of the Hamiltonian in Eqs.~(\ref{sl625_ham1}) and (\ref{sl625_ham2}) obtained in the PSG representation with $\{\eta_\T,\eta,\eta_1,\eta_2,\eta_z\}=\{1,1,1,1,1\}$ and $\{g_\T,g_{P_x},g_{P_y},g_{P_{xy}},g_{P_z}\}=\{i\tau^2,\tau^0,\tau^0,\tau^0,i\tau^3\}$. Panel (a) shows the bands for a system with torus geometry (i.e., periodic boundary conditions in both directions), while (b) depicts the bands for a cylinder geometry (the cylinder axis is along the $y$-direction). The upper and lower blocks in Eq.~(\ref{sl625_ham1}) yield degenerate bulk bands thoughout the Brillouin zone. The spectrum is similar to the Bernevig-Hughes-Zhang (BHZ) model describing the bands of HgTe: While the bulk is gapped, a pair of counter-propagating gapless states appears at the boundary of the system. In contrast to the BHZ model, however, the present system is not a topological insulator but a spinon version of a time-reversal invariant topological superconductor with Majorana edge modes. Note that (b) only shows the edge states at one cylinder edge, where red (green) lines correspond to $\uparrow$ ($\downarrow$) states. The boundary modes at the other edge have the same dispersion but with the spin orientations reversed.}
\label{fig:psg625}
\end{figure}

The Hamiltonian in Eqs.~(\ref{sl625_ham1}) and (\ref{sl625_ham2}) resembles the Bernevig-Hughes-Zhang (BHZ) model\cite{bernevig06,konig07} which describes the band structure of the topological insulator material HgTe. Particularly, the two $2\times 2$ blocks of the Hamiltonian represent time-reversal versions of each other. From the Chern numbers of the two blocks, $n_\uparrow=1$ and $n_\downarrow=-1$, one finds a non-trivial $\mathds{Z}_2$ topological invariant $\nu=1$ [see Eq.~(\ref{z2_invariant})], indicating that the system resides in a topological phase. The band structure in Fig.~\ref{fig:psg625} indeed shows a bulk gap and a pair of counter-propagating gapless edge states. While the form of the Hamiltonian is similar to the BHZ model, its interpretation is rather different. Most importantly, in contrast to the BHZ model, the components of $\hat{\Psi}_\rr=(f_{\rr\uparrow},f_{\rr\uparrow}^\dagger,f_{\rr\downarrow},f_{\rr\downarrow}^\dagger)^\text{T}$ are related by Hermitian conjugation such that spinon modes with positive and negative energies are not distinct quantum states. Correspondingly, the sin terms describe $p$-wave pairing for the spinons instead of spin-orbit coupling, and the edge states are Majorana modes. The system may, hence, be considered as two time-reversed copies of a $p_x\pm ip_y$ `spinon superconductor' with different chiralities [we note in passing that a spinon version of a topological insulator is unstable in two dimensions due to U(1) gauge fluctuations, see e.g. Ref.~\onlinecite{krempa10}].

The non-trivial $\mathds{Z}_2$ topological invariant implies that the gaplessness of the edge states is protected by time-reversal symmetry. Denoting the right moving (spin up) Majorana modes by $\gamma_\text{R}$ and the left moving (spin down) Majorana modes by $\gamma_\text{L}$, time reversal acts as $\T\gamma_\text{R}\T^\dagger=\gamma_\text{L}$ and $\T\gamma_\text{L}\T^\dagger=-\gamma_\text{R}$. Hence, the only mass term $i\gamma_\text{R}\gamma_\text{L}$ that can gap out the edge states changes sign under $\T$ and is therefore forbidden by time-reversal symmetry. The non-trivial action of $P_z$ has further consequences on the topological protection of the edge states. Since $g_{P_z}=i\tau^3$, the spin down edge mode $\gamma_\text{L}$ picks up a minus sign under inversion, i.e., $P_z\gamma_\text{R}P_z^\dagger=\gamma_\text{R}$ and $P_z\gamma_\text{L}P_z^\dagger=-\gamma_\text{L}$ (this minus sign would not occur within a trivial implementation of $P_z$). As a result, the term $i\gamma_\text{R}\gamma_\text{L}$ is also forbidden by inversion $P_z$ such that gapping out the edge states requires the violation of $\T$ and $P_z$ {\it simultaneously}. Since there is no perturbation gapping out the boundary modes by {\it only} violating $\T$, the topological protection within this PSG representation goes beyond the protection of an ordinary time-reversal invariant topological superconductor.
%
%%%%%%%%%%%%%%%%%%%%%%%%%%% Figure 4 %%%%%%%%%%%%%%%%%%%%%%%%%%%
%
\begin{figure}[t]
\centering
\includegraphics[width=0.99\linewidth]{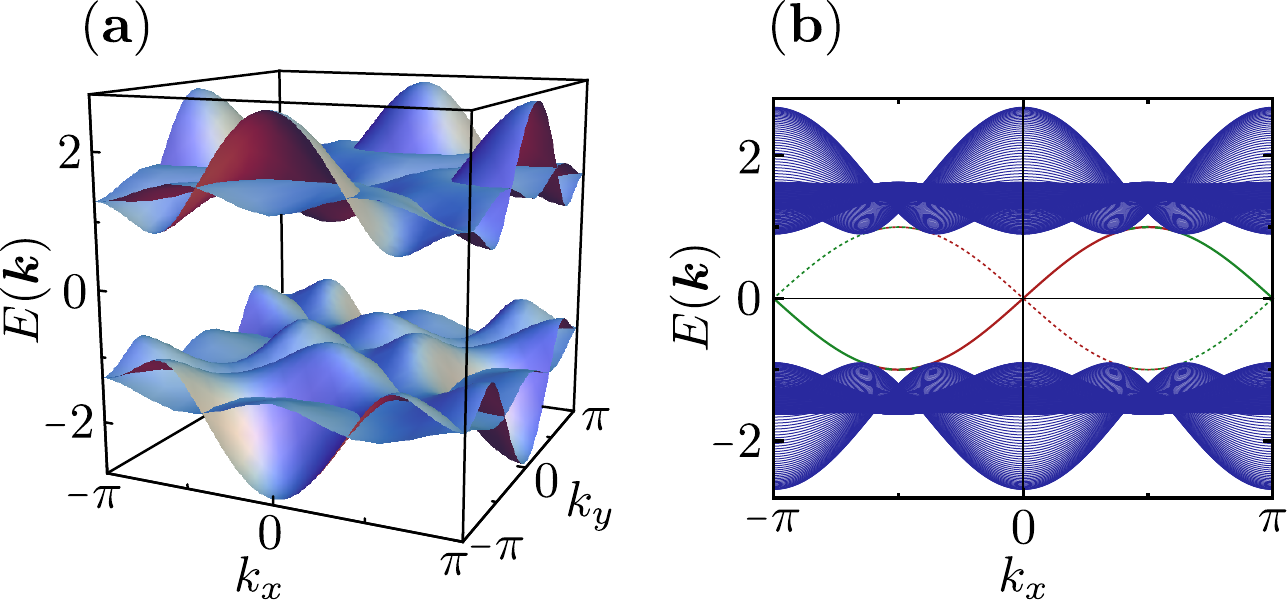}
\caption{Spectrum of the Hamiltonian in Eqs.~(\ref{sl696_ham1}) and (\ref{sl696_ham2}) obtained in the PSG representation with $\{\eta_\T,\eta,\eta_1,\eta_2,\eta_z\}=\{-1,1,-1,-1,1\}$ and $\{g_\T,g_{P_x},g_{P_y},g_{P_{xy}},g_{P_z}\}=\{i\tau^2,i\tau^3,i\tau^3,\tau^0,i\tau^3\}$. Panel (a) depicts the band structure for the system on a torus. In contrast to Fig.~\ref{fig:psg625}, the bulk bands of the upper and lower blocks in Eq.~(\ref{sl696_ham1}) are not degenerate in the whole Brillouin zone. Panel (b) displays the bands for cylinder geometry showing a pair of counter-propagating edge states where red (green) corresponds to $\uparrow$ ($\downarrow$) modes. Since time reversal is now also associated with an overall momentum transfer, the spin down modes are shifted by $\pi$ in $k_x$-direction [compare to Fig.~\ref{fig:psg625} (b)]. Full lines illustrate the edge states on one cylinder edge, while dashed lines denote the boundary modes on the the opposite edge (with the same color coding for the spin directions).}
\label{fig:psg696}
\end{figure}
%

%%%%%%%%%%%%%%%%%%%%%%%%%%%%%%%%%%%%%%%%%%%%%%%%%%%%%%%%%%%%%%
% V.B. EXAMPLE 2
%%%%%%%%%%%%%%%%%%%%%%%%%%%%%%%%%%%%%%%%%%%%%%%%%%%%%%%%%%%%%%

\subsection{PSG with $\{\eta_\T,\eta,\eta_1,\eta_2,\eta_z\}=\{-1,1,-1,-1,1\}$ and $\{g_\T,g_{P_x},g_{P_y},g_{P_{xy}},g_{P_z}\}=\{i\tau^2,i\tau^3,i\tau^3,\tau^0,i\tau^3\}$}\label{sl696}
Next we discuss a PSG representation where projective time reversal differs from the trivial one. As compared to time reversal in an ordinary fermion system where $g_\T=i\tau^2$ and $\eta_\T=1$, here we have $\eta_\T=-1$. According to Eq.~(\ref{g_solutions}) this leads to an additional site dependent factor $(-1)^{x+y}$ in the action of $\T$. Specifically, when applied to the spinor $\hat{\Psi}_\rr$, time reversal yields
\begin{equation}
\T\hat{\Psi}_\rr \T^\dagger=(-1)^{x+y}
\left(\begin{array}{cc}
0 & \tau^0 \\
-\tau^0 & 0
\end{array}\right)\hat{\Psi}_\rr\;.
\end{equation}
Similarly, in $k$-space one obtains
\begin{equation}
\T\hat{\Psi}_\kk \T^\dagger=
\left(\begin{array}{cc}
0 & \tau^0 \\
-\tau^0 & 0
\end{array}\right)\hat{\Psi}_{-\kk+(\pi,\pi)}\;,
\end{equation}
where in addition to momentum inversion $\kk\rightarrow -\kk$, time reversal now also {\it shifts} the momentum by $(\pi,\pi)$. Most importantly, however, $\T$ is still an antiunitary operator with $\T^2|\phi\rangle=-|\phi\rangle$, implying that Kramer's theorem as well as the $\mathds{Z}_2$ topological classification still apply. This can be seen using the arguments from Fig.~\ref{fig:z2_invariant}. Since $\T$ comes along with a momentum transformation $\kk\rightarrow -\kk+(\pi,\pi)$, there are again two time-reversal invariant momenta for a system on a cylinder, now given by $k_x=-\pi/2$ and $k_x=\pi/2$. At these momenta all states must at least be two-fold degenerate. Apart from the fact that the time-reversal invariant momenta are shifted by $\pi/2$ as compared to Fig.~\ref{fig:z2_invariant}, the above discussion still applies, showing that there are two topologically distinct band structures. This shift does not effect the computation of the $\mathds{Z}_2$ topological invariant such that Eq.~(\ref{z2_invariant}) is still valid.

In order to discuss a specific example, we consider the following non-vanishing mean-field parameters,
\begin{eqnarray}
&t_{1,\drr=(1,0)}^0=-t_{1,\drr=(0,1)}^0=-1\;,&\notag\\
&t_{2,\drr=(1,0)}^2=-t_{3,\drr=(0,1)}^2=-1\;,&\notag\\
&s_{\drr=(1,1)}^3=s_{\drr=(1,-1)}^3=0.5\;.&
\end{eqnarray}
In the basis of $\hat{\Psi}_\kk$ the Hamiltonian has the form 
\begin{equation}
H_\text{mf}=\frac{1}{2}\sum_\kk\hat{\Psi}_\kk^\dagger
\left[\left(\begin{array}{cc}
h_\kk & 0 \\
0 & h^*_{-\kk+(\pi,\pi)}
\end{array}\right)+a_3
\left(\begin{array}{cc}
\tau^3 & 0 \\
0 & \tau^3
\end{array}\right)\right]
\hat{\Psi}_\kk\label{sl696_ham1}
\end{equation}
with
\begin{widetext}
\begin{equation}
h_\kk=
\left(\begin{array}{cc}
-\cos(k_x)+\cos(k_y)+\cos(k_x)\cos(k_y) & \sin(k_x)-i\sin(k_y)\\
\sin(k_x)+i\sin(k_y) & \cos(k_x)-\cos(k_y)-\cos(k_x)\cos(k_y)
\end{array}\right)\label{sl696_ham2}
\end{equation}
\end{widetext}
The Lagrange multipliers are calculated numerically, yielding $a_1=a_2=0$ and $a_3\approx 0.607$. The spin up and spin down blocks in the Hamiltonian again decouple (this may always be achieved within the present PSG representation using an appropriate gauge choice). Furthermore, the two blocks are time-reversed analogues of each other. The crucial difference as compared to Eq.~(\ref{sl625_ham2}) is that the kinetic $\cos(k_x)$ and $\cos(k_y)$ terms now have opposite signs.

The Chern numbers of the two blocks are $n_\uparrow=-n_\downarrow=1$, leading to a non-trivial $\mathds{Z}_2$ topological invariant $\nu=1$. Accordingly, the band structure in Fig.~\ref{fig:psg696} shows a pair of counter-propagating Majorana edge states for each end of the cylinder. When compared to Fig.~\ref{fig:psg625}, the spin down boundary modes are shifted by $\pi$ in $k_x$-direction, which is a consequence of the additional momentum transfer associated with $\T$. In this PSG representation, the two boundary modes at each cylinder cap, denoted by $\gamma_\text{R}$ and $\gamma_\text{L}$, (where the right-mover $\gamma_\text{R}$ carries spin up while the left-mover $\gamma_\text{L}$ carries spin down) enjoy an exceptionally strong protection. Firstly, the edge states are protected by time reversal which again acts as $\T\gamma_\text{R}\T^\dagger=\gamma_\text{L}$, $\T\gamma_\text{L}\T^\dagger=-\gamma_\text{R}$ and therefore forbids the mass term $i\gamma_\text{R}\gamma_\text{L}$. Furthermore, since a coupling between $\gamma_\text{R}$ and $\gamma_\text{L}$ requires a momentum transfer of $\Delta k_x=\pi$, the term $i\gamma_\text{R}\gamma_\text{L}$ necessarily violates translation invariance $T_x$ (but preserves $T_x^2$). As in Sec.~\ref{sl625}, inversion $P_z$ is implemented by $\eta_z=1$, $g_{P_z}=i\tau^3$. With the arguments given above, it follows that $i\gamma_\text{R}\gamma_\text{L}$ is odd under $P_z$. Finally, if the edges are along the $x$-direction, the system fulfills mirror symmetry $P_x$ on the cylinder. One can show that in this case the term $i\gamma_\text{R}\gamma_\text{L}$ is also forbidden by $P_x$. In total, the Hamiltonian in Eqs.~(\ref{sl696_ham1}) and (\ref{sl696_ham2}) has the remarkable property that $\T$, $T_x$, $P_x$ and $P_z$ need to be broken {\it simultaneously} in order to gap out the edge states at zero energy. Since the gaplessness of the boundary modes does not only rely on $\T$, this system resembles a topological crystalline insulator.

%%%%%%%%%%%%%%%%%%%%%%%%%%%%%%%%%%%%%%%%%%%%%%%%%%%%%%%%%%%%%%
% V.C. EXAMPLE 3
%%%%%%%%%%%%%%%%%%%%%%%%%%%%%%%%%%%%%%%%%%%%%%%%%%%%%%%%%%%%%%

\subsection{PSG with $\{\eta_\T,\eta,\eta_1,\eta_2,\eta_z\}=\{-1,1,-1,-1,1\}$ and $\{g_\T,g_{P_x},g_{P_y},g_{P_{xy}},g_{P_z}\}=\{\tau^0,i\tau^1,i\tau^1,i\tau^3,i\tau^3\}$}\label{sl504}
In the next example, we consider a PSG representation with $g_\T=\tau^0$. Such an implementation of time reversal is fundamentally different from the cases studied above, since no gauge transformation can convert $g_\T=\tau^0$ into $g_\T=i\tau^2$. The action on $\hat{\Psi}_\rr$ is now given by
\begin{equation}
\T\hat{\Psi}_\rr \T^\dagger=(-1)^{x+y}
\left(\begin{array}{cc}
\tau^0 & 0 \\
0 & \tau^0
\end{array}\right)\hat{\Psi}_\rr\;,
\end{equation}
implying that $\T^2|\phi\rangle=|\phi\rangle$. Thus, the requirements for Kramer's theorem are not fulfilled and a $\mathds{Z}_2$ topological invariant (at least based on time reversal) does not exist. In other words, the system falls into the class BDI (see Refs.~\onlinecite{hasan10} and \onlinecite{kitaev09}) which does not have a topological index (in contrast, the examples studied before belong to class DIII). A topological protection of edge states is generally not given in this PSG representation.

\begin{figure}[t]
\centering
\includegraphics[width=0.99\linewidth]{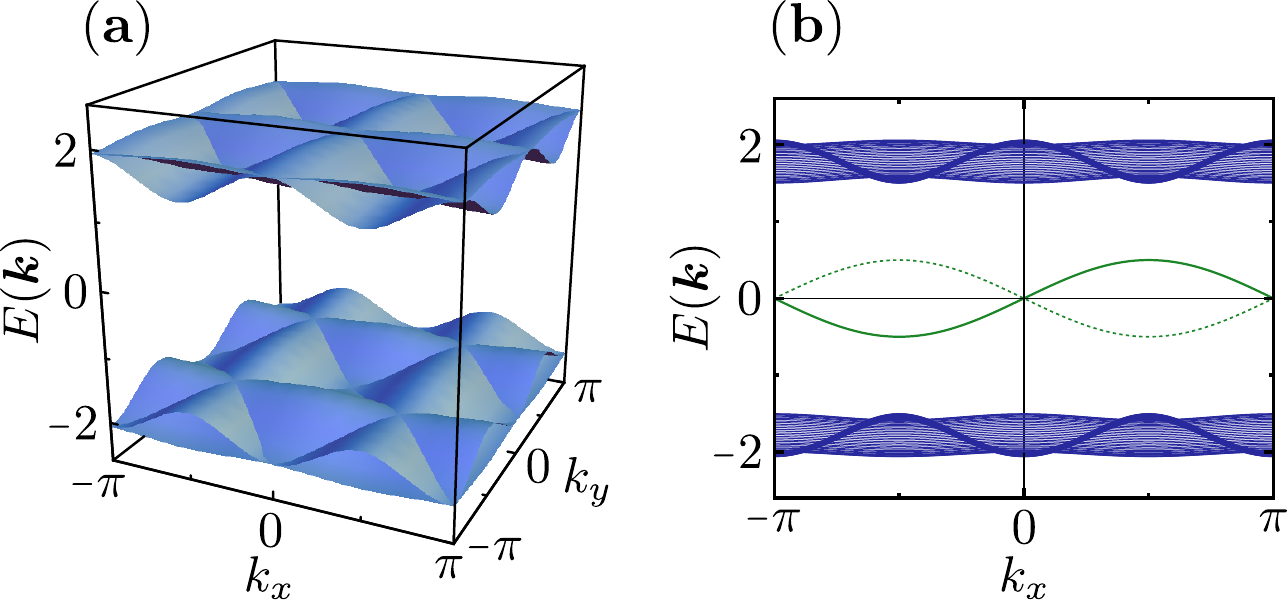}
\caption{Example for a band structure in the PSG representation with $\{\eta_\T,\eta,\eta_1,\eta_2,\eta_z\}=\{-1,1,-1,-1,1\}$ and $\{g_\T,g_{P_x},g_{P_y},g_{P_{xy}},g_{P_z}\}=\{\tau^0,i\tau^1,i\tau^1,i\tau^3,i\tau^3\}$ using the parameters in Eq.~(\ref{sl504_parameters}). As shown in (a), the system on a torus features a bulk spinon gap. Considering a cylinder geometry [see (b)], the model exhibits an edge state; however, due to the absence of a topological index, such modes are not topologically protected (there is always an even number of level crossings at fixed energy). Interestingly, boundary modes only carry spin down. The dashed green line refers to bound states on the opposite edge.}
\label{fig:psg504}
\end{figure}

We again consider a specific example with non-trivial mean-field amplitudes
\begin{eqnarray}
&s_{\drr=(1,0)}^3=s_{\drr=(0,1)}^3=1\;,&\notag\\
&t_{1,\drr=(1,0)}^0=-t_{1,\drr=(0,1)}^0=1\;,&\notag\\
&t_{2,\drr=(1,0)}^2=t_{3,\drr=(0,1)}^2=1\;,&\notag\\
&t_{3,\drr=(1,0)}^1=t_{2,\drr=(0,1)}^1=-0.5\;.&\label{sl504_parameters}
\end{eqnarray}
In $k$-space the Hamiltonian reads
\begin{equation}
H_\text{mf}=\frac{1}{2}\sum_\kk\hat{\Psi}_\kk^\dagger
\left(\begin{array}{cc}
h^1_\kk & 0 \\
0 & h^2_\kk
\end{array}\right)
\hat{\Psi}_\kk\label{sl504_ham1}
\end{equation}
with
\begin{equation}
h^1_\kk=
\left(\begin{array}{cc}
2\cos(k_x) & -\frac{3}{2}\sin(k_x)-\frac{i}{2}\sin(k_y)\\
-\frac{3}{2}\sin(k_x)+\frac{i}{2}\sin(k_y) &-2\cos(k_x)
\end{array}\right)\label{sl504_ham2}
\end{equation}
\begin{equation}
h^2_\kk=
\left(\begin{array}{cc}
2\cos(k_y) & -\frac{1}{2}\sin(k_x)+\frac{3i}{2}\sin(k_y)\\
-\frac{1}{2}\sin(k_x)-\frac{3i}{2}\sin(k_y) &-2\cos(k_y)
\end{array}\right)\;,\label{sl504_ham3}
\end{equation}
where all Lagrange multipliers vanish. Within an appropriate gauge choice, the Hamiltonian is always block-diagonal. Since $\T$ does not couple spin up and spin down sectors, there is no simple relation between the two blocks in the Hamiltonian. As illustrated in Fig.~\ref{fig:psg504}, edge states may still exist; however, they are not connected to the bulk and could, in principle, be removed without changing the topology of the system. The band structure is, hence, topologically equivalent to that of a trivial superconductor.
%
%%%%%%%%%%%%%%%%%%%%%%%%%%% Figure 5 %%%%%%%%%%%%%%%%%%%%%%%%%%%
%
%
%%%%%%%%%%%%%%%%%%%%%%%%%%% Table 1 %%%%%%%%%%%%%%%%%%%%%%%%%%%
%
\begin{table}[t]
\centering
\begin{tabular}{|c||c|c|c|c|}
\hline
\rule{0pt}{0.4cm}            & $i\gamma_\text{R}\gamma_\text{L}$ & $i\eta_\text{R}\eta_\text{L}$                 & $i\gamma_\text{R}\eta_\text{L}$ &  $i\gamma_\text{L}\eta_\text{R}$ \\[0.15cm]
\hline\hline
\rule{0pt}{0.4cm}$\T$      & $-$                                                            & $-$                                                            & $i\gamma_\text{L}\eta_\text{R}$  & $i\gamma_\text{R}\eta_\text{L}$\\[0.15cm]
\hline
\rule{0pt}{0.4cm}$T_x$  & $+$                                                           & $+$                                                           & $-$                                                    & $-$\\[0.2cm]
\hline
\rule{0pt}{0.4cm}$P_x$  & $+$                                                           & $+$                                                           & $i\gamma_\text{L}\eta_\text{R}$  & $i\gamma_\text{R}\eta_\text{L}$\\[0.15cm]
\hline
\rule{0pt}{0.4cm}$P_z$   & $i\eta_\text{R}\eta_\text{L}$                 & $i\gamma_\text{R}\gamma_\text{L}$ &  $i\gamma_\text{L}\eta_\text{R}$ & $i\gamma_\text{R}\eta_\text{L}$\\[0.15cm]
\hline
\end{tabular}
\caption{Transformation of the terms $i\gamma_\text{R}\gamma_\text{L}$, $i\eta_\text{R}\eta_\text{L}$, $i\gamma_\text{R}\eta_\text{L}$, and $i\gamma_\text{L}\eta_\text{R}$ under the effect of $\T$, $T_x$, $P_x$, and $P_z$. The Majorana modes $\gamma_\text{R}$, $\gamma_\text{L}$, $\eta_\text{R}$, and $\eta_\text{L}$ represent the edge states of the PSG representation in Sec.~\ref{sl81} at zero energy. Note that an entry ``$+$" (``$-$") indicates that under the respective symmetry, the term remains invariant (changes sign).}
\label{psg81_table}
\end{table}
%

%%%%%%%%%%%%%%%%%%%%%%%%%%%%%%%%%%%%%%%%%%%%%%%%%%%%%%%%%%%%%%
% V.D. EXAMPLE 4
%%%%%%%%%%%%%%%%%%%%%%%%%%%%%%%%%%%%%%%%%%%%%%%%%%%%%%%%%%%%%%

\subsection{PSG with $\{\eta_\T,\eta,\eta_1,\eta_2,\eta_z\}=\{1,1,1,1,-1\}$ and $\{g_\T,g_{P_x},g_{P_y},g_{P_{xy}},g_{P_z}\}=\{i\tau^2,\tau^0,\tau^0,\tau^0,\tau^0\}$}\label{sl81}
We continue with a spin liquid phase that exhibits a more complex edge-state structure. In this PSG representation, projective time reversal and trivial time reversal are identical such that a $\mathds{Z}_2$ topological invariant exists. The implementation of symmetries is similar to the PSG representation in Sec.~\ref{sl625}, but with a different action of $P_z$. Here, inversion $P_z$ has a real-space dependence resulting from $\eta_z=-1$. An instructive example is given by the mean-field parameters 
\begin{eqnarray}
&t_{2,\drr=(1,0)}^1=-t_{3,\drr=(0,1)}^1=1\;,&\notag\\
&s_{\drr=(1,1)}^3=s_{\drr=(1,-1)}^3=1\;.&\label{sl81_parameters}
\end{eqnarray}
The corresponding Hamiltonian in $k$-space becomes
\begin{equation}
H_\text{mf}=\frac{1}{2}\sum_\kk\hat{\Psi}_\kk^\dagger
\left(\begin{array}{cc}
h_\kk & 0 \\
0 & h^*_{-\kk}
\end{array}\right)
\hat{\Psi}_\kk\;,\label{sl81_ham1}
\end{equation}
where all Lagrange multipliers vanish.\cite{lagrange} Furthermore, one obtains
\begin{equation}
h_\kk=
\left(\begin{array}{cc}
2\cos(k_x)\cos(k_y) & -i\sin(k_x)-\sin(k_y)\\
i\sin(k_x)-\sin(k_y) & -2\cos(k_x)\cos(k_y)
\end{array}\right)\;.\label{sl81_ham2}
\end{equation}
Note that in this PSG representation the spin up and spin down sectors are not necessarily decoupled. Here, however, for simplicity we choose an example where the Hamiltonian is block diagonal. As compared to Eq.~(\ref{sl625_ham2}) the present model does not exhibit kinetic $\cos(k_x)+\cos(k_y)$ terms. We emphasize that by setting $s_{\drr=(1,0)}^3=s_{\drr=(0,1)}^3=0$ and $s_{\drr=(1,1)}^3=s_{\drr=(1,-1)}^3=1$ in Eq.~(\ref{sl625_parameters}), the two Hamiltonians are identical, demonstrating that a spin phase with the same spinon edge-state structure may also occur in the PSG representation of Sec.~\ref{sl625}. Indeed, similar band structures occur in many PSG representations and represent a generic example of spinon spectra in the case of broken spin-rotation symmetries. 
%
%%%%%%%%%%%%%%%%%%%%%%%%%%% Figure 6 %%%%%%%%%%%%%%%%%%%%%%%%%%%
%
\begin{figure}[t]
\centering
\includegraphics[width=0.99\linewidth]{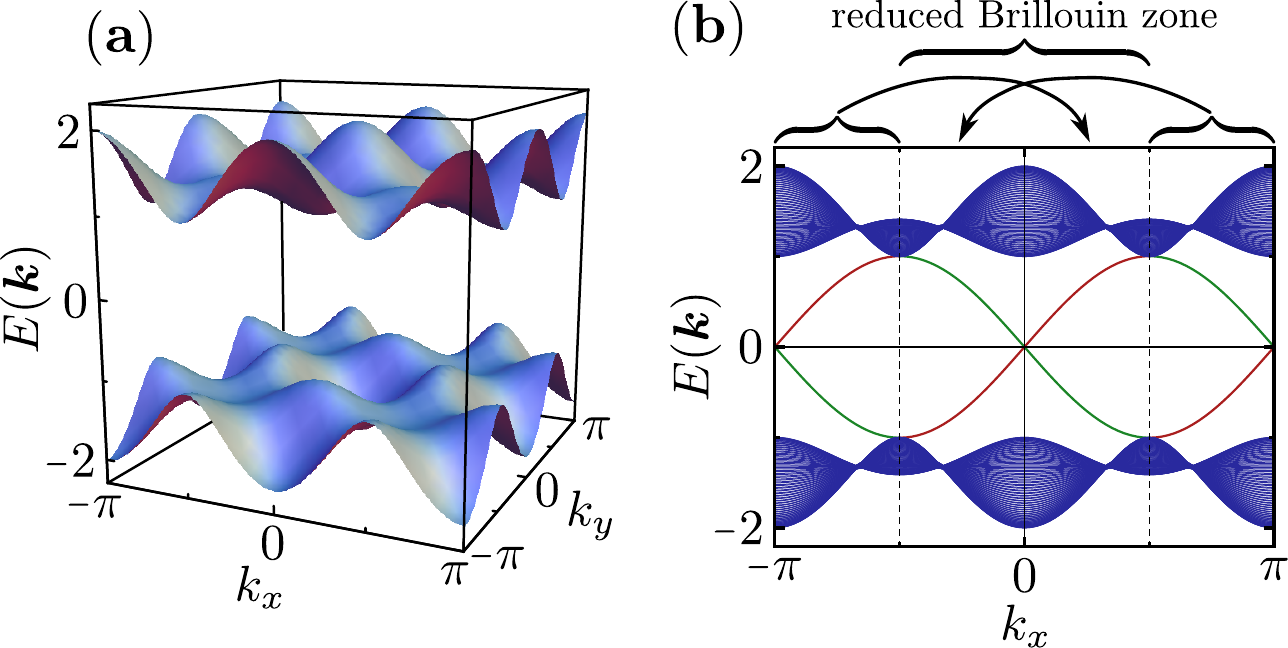}
\caption{Band structure in the PSG representation with $\{\eta_\T,\eta,\eta_1,\eta_2,\eta_z\}=\{1,1,1,1,-1\}$ and $\{g_\T,g_{P_x},g_{P_y},g_{P_{xy}},g_{P_z}\}=\{i\tau^2,\tau^0,\tau^0,\tau^0,\tau^0\}$ for a system (a) on a torus and (b) on a cylinder, using the parameters in Eq.~(\ref{sl81_parameters}). As shown in (b), the edge-state spectrum consists of two right-moving spin up modes (red) and two left-moving spin down modes (green). Even though the $\mathds{Z}_2$ topological invariant is trivial, $\nu=0$, one may still assign a modified {\it non-trivial} $\mathds{Z}_2$ index $\nu'=1$ to this band structure, see text for details. (b) also illustrates the back folding of bands when doubling the unit cell in $x$-direction.}
\label{fig:psg81}
\end{figure}

The Chern numbers of the two blocks in Eq.~(\ref{sl81_ham1}) are $n_\uparrow=-n_\downarrow=2$, resulting in two right-moving and two left-moving modes; see the band structure in Fig.~\ref{fig:psg81}. Pairs of modes with opposite chirality are located at $k_x=0$ and at $k_x=\pi$, respectively. According to Eq.~(\ref{z2_invariant}) the $\mathds{Z}_2$ index is $\nu=0$, and one would expect the system to be topologically trivial. However, the spectrum never shows a gap in the edge states, even if arbitrary symmetry allowed terms are added to the Hamiltonian. We will argue that  this gaplessness indeed follows from the fact that -- based on a slightly modified $\mathds{Z}_2$ index -- the system can be classified as topologically {\it non-trivial}, even though $\nu=0$.

To understand this peculiarity, we study the symmetry properties of the Majorana modes at $E=0$, where the right-moving and left-moving modes at $k_x=0$ ($k_x=\pi$) are denoted by $\gamma_\text{R}$ and $\gamma_\text{L}$ ($\eta_\text{R}$ and $\eta_\text{L}$), respectively. It will prove convenient to change the real-space description of the system by considering an enlarged unit cell that extends over {\it two} sites in the $x$-direction and one site in the $y$-direction. (We emphasize, however, that this unit cell expansion is cosmetic and does not change the terms in the Hamiltonian. Particularly, we do not break translation invariance $T_x$.) In this new description the first Brillouin zone for a system on a cylinder shrinks down to $k_x\in[-\pi/2,\pi/2]$, while the total number of bands is doubled. Specifically, each momentum $k_x\in[-\pi/2,\pi/2]$ of the reduced Brillouin zone now labels the original states at this $k_x$ as well as the states with initial momentum $k_x+\pi$ [this back folding is also illustrated in Fig.~\ref{fig:psg81} (b)]. Within the PSG representation considered here, the new Brillouin zone contains two copies of degenerate band structures (in the following referred to as sets): the original bands at $k_x\in[-\pi/2,\pi/2]$ and the shifted ones from $k_x\in[\pi/2,3\pi/2]$. Note that all zero-energy boundary modes $\gamma_\text{R}$, $\gamma_\text{L}$, $\eta_\text{R}$, and $\eta_\text{L}$ now carry $k_x=0$. Couplings between such modes are described by the four terms $i\gamma_\text{R}\gamma_\text{L}$, $i\eta_\text{R}\eta_\text{L}$, $i\gamma_\text{R}\eta_\text{L}$, and $i\gamma_\text{L}\eta_\text{R}$. By explicitly calculating the edge-state wave functions, one finds that these terms transform under $\T$, $T_x$, $P_x$, and $P_z$ according to Table~\ref{psg81_table} (the table only includes the symmetries of a cylinder in the $y$-direction, i.e., $T_y$ and $P_{xy}$ are ignored).

First, we only consider coupling between $\gamma_\text{R}$ and $\gamma_\text{L}$, as well as between $\eta_\text{R}$ and $\eta_\text{L}$, i.e., we neglect the last two columns of Table~\ref{psg81_table}. The reduced Brillouin zone is periodic with respect to $\Delta k_x=\pi$, such that the momenta $k_x=-\pi/2$ and $k_x=\pi/2$ are equivalent. Hence, the new Brillouin zone still contains {\it two} time reversal invariant momenta at $k_x=0$ and $k_x=\pi/2$. It follows that the classification of topologically distinct band structures shown in Fig.~\ref{fig:z2_invariant} is still valid in the reduced zone scheme, which allows us to define a modified $\mathds{Z}_2$ topological invariant $\nu'$,
\begin{equation}
\nu'=\frac{n'_\uparrow-n'_\downarrow}{2}\;\text{mod}\;2\;,\label{z2_invariant2}
\end{equation}
where the Chern numbers $n'_\uparrow$ and $n'_\downarrow$ are only calculated in the new Brillouin zone. For both sets of degenerate bands (i.e., those originally located at $k_x=[-\pi/2,\pi/2]$ and the shifted ones from $k_x\in[\pi/2,3\pi/2]$) we find $n'_\downarrow=-n'_\uparrow=1$ yielding $\nu'=1$. This result indicates that if both sets are considered independently, their bands exhibit a non-trivial topology in the reduced zone scheme. Indeed, Table~\ref{psg81_table} shows that the edge states within each set are separately protected by time reversal symmetry: Ignoring the coupling between $\gamma_\text{R/L}$ and $\eta_\text{R/L}$, there exists a term $i\gamma_\text{R}\gamma_\text{L}+i\eta_\text{R}\eta_{L}$ which {\it only} breaks $\T$ and gaps out the edge states. On the other hand, none of the symmetries $T_x$, $P_x$, $P_z$ can be violated, without also breaking $\T$.

We now consider coupling between $\gamma_\text{R/L}$ and $\eta_\text{R/L}$. Most importantly, according to Table~\ref{psg81_table}, there is no linear combination of $i\gamma_\text{R}\eta_\text{L}$ and $i\gamma_\text{L}\eta_\text{R}$ that preserves all symmetries of the system. Any coupling between the two sets is forbidden, implying that $\gamma_\text{R/L}$ and $\eta_\text{R/L}$ may indeed be treated independently. The non-trivial index $\nu'=1$, consequently, still yields a valid classification. Specifically, Table~\ref{psg81_table} shows that a coupling among both sets is forbidden by translation invariance $T_x$: There exists a term $i\gamma_\text{R}\eta_\text{L}+i\gamma_\text{L}\eta_\text{R}$ gapping out the edge states, which {\it only} breaks $T_x$ but no other symmetries. On the other hand, there is no linear combination of $i\gamma_\text{R}\eta_\text{L}$ and $i\gamma_\text{L}\eta_\text{R}$ which breaks any of the symmetries $\T$, $P_x$ or $P_z$ but respects $T_x$. This is intuitively clear, since any term coupling $\gamma_\text{R/L}$ and $\eta_\text{R/L}$ must be associated with a momentum transfer of $\Delta k_x=\pi$ in the original Brillouin zone, which is not possible without breaking $T_x$.

In total, we have shown that the edge states are topologically protected by $\T$ and $T_x$, where violating one of them already gaps out the system entirely. Due to time reversal $\T$, there exists a modified $\mathds{Z}_2$ topological invariant $\nu'$ (which is non-trivial in the case studied here), defined for both sets of bands in the reduced Brillouin zone. Furthermore, as a result of translation invariance $T_x$, coupling between the sets is forbidden and the topological index survives.\cite{z2_index}

In this discussion, we have assumed that the system on a cylinder fulfills mirror symmetry $P_x$, which implies that the edge is along the $x$-direction. For an arbitrary cylinder edge, $P_x$ is already violated by the underlying geometry, such that $P_x$ must be excluded from the considerations above. This, however, does not change our conclusions about the edge-state protection.
%
%%%%%%%%%%%%%%%%%%%%%%%%%%% Table 2 %%%%%%%%%%%%%%%%%%%%%%%%%%%
%
\begin{table}[t]
\centering
\begin{tabular}{|c||c|c|c|c|}
\hline
\rule{0pt}{0.4cm}            & $i\gamma_\text{R}\gamma_\text{L}$ & $i\eta_\text{R}\eta_\text{L}$                  & $i\gamma_\text{R}\eta_\text{L}$ &  $i\gamma_\text{L}\eta_\text{R}$ \\[0.15cm]
\hline\hline
\rule{0pt}{0.4cm}$\T$     & $-i\eta_\text{R}\eta_\text{L}$               & $-i\gamma_\text{R}\gamma_\text{L}$ & $-$                                                     & $-$\\[0.15cm]
\hline
\rule{0pt}{0.4cm}$T_x$  & $+$                                                           & $+$                                                           & $-$                                                    & $-$\\[0.2cm]
\hline
\rule{0pt}{0.4cm}$P_x$  & $-$                                                           & $-$                                                           & $i\gamma_\text{L}\eta_\text{R}$  & $i\gamma_\text{R}\eta_\text{L}$\\[0.15cm]
\hline
\rule{0pt}{0.4cm}$P_z$   & $-i\eta_\text{R}\eta_\text{L}$              & $-i\gamma_\text{R}\gamma_\text{L}$ &  $-i\gamma_\text{L}\eta_\text{R}$ & $-i\gamma_\text{R}\eta_\text{L}$\\[0.15cm]
\hline
\end{tabular}
\caption{Symmetry properties of the terms $i\gamma_\text{R}\gamma_\text{L}$, $i\eta_\text{R}\eta_\text{L}$, $i\gamma_\text{R}\eta_\text{L}$, and $i\gamma_\text{L}\eta_\text{R}$ under the effect of $\T$, $T_x$, $P_x$, and $P_z$. The Majorana modes $\gamma_\text{R}$, $\gamma_\text{L}$, $\eta_\text{R}$, and $\eta_\text{L}$ represent the edge states of the PSG representation in Sec.~\ref{sl101} at zero energy. Note that an entry ``$+$" (``$-$") indicates that under the respective symmetry, the term remains invariant (changes sign).}
\label{psg101_table}
\end{table}
%

%%%%%%%%%%%%%%%%%%%%%%%%%%%%%%%%%%%%%%%%%%%%%%%%%%%%%%%%%%%%%%
% V.E. EXAMPLE 5
%%%%%%%%%%%%%%%%%%%%%%%%%%%%%%%%%%%%%%%%%%%%%%%%%%%%%%%%%%%%%%

\subsection{PSG with $\{\eta_\T,\eta,\eta_1,\eta_2,\eta_z\}=\{-1,1,1,1,-1\}$ and $\{g_\T,g_{P_x},g_{P_y},g_{P_{xy}},g_{P_z}\}=\{i\tau^2,i\tau^3,i\tau^3,\tau^0,\tau^0\}$}\label{sl101}
We now discuss a PSG representation which -- regarding the form of its mean-field ans\"atze -- is similar to the one studied in Sec.~\ref{sl81}. In contrast to the previous section, however, the symmetries act in a very different way. Time reversal is implemented via $\eta_\T=-1$ and $g_\T=i\tau^2$ and, therefore, has a real-space dependence. As argued in Sec.~\ref{sl696}, Kramer's theorem still holds, leading to a $\mathds{Z}_2$ topological classification. Using the mean-field parameters
\begin{eqnarray}
&t_{3,\drr=(1,0)}^2=-t_{2,\drr=(0,1)}^2=1\;,&\notag\\
&s_{\drr=(1,1)}^3=s_{\drr=(1,-1)}^3=-1\;,&
\end{eqnarray}
the Hamiltonian in $k$-space becomes
\begin{equation}
H_\text{mf}=\frac{1}{2}\sum_\kk\hat{\Psi}_\kk^\dagger
\left(\begin{array}{cc}
h_\kk & 0 \\
0 & h^*_{-\kk+(\pi,\pi)}
\end{array}\right)
\hat{\Psi}_\kk\label{sl101_ham1}
\end{equation}
where
\begin{equation}
h_\kk=
\left(\begin{array}{cc}
-2\cos(k_x)\cos(k_y) & -i\sin(k_x)+\sin(k_y)\\
i\sin(k_x)+\sin(k_y) & 2\cos(k_x)\cos(k_y)
\end{array}\right)\;.\label{sl101_ham2}
\end{equation}
Note that all Lagrange multipliers vanish. The terms in Eq.~(\ref{sl101_ham2}) differ from those in Eq.~(\ref{sl81_ham2}) only by signs. Diagonalizing the Hamiltonian, it even turns out that the band structure is identical (see Fig.~\ref{fig:psg81}), with Chern numbers again given by $n_\uparrow=-n_\downarrow=2$. We repeat the procedure of the previous section, showing that even though $\nu=0$, the system is still topologically non-trivial. Applying the arguments from above, we study the system within an enlarged unit cell. Given that the new Brillouin zone has a periodicity of $\Delta k_x=\pi$ and that time reversal $\T$ is associated with a $k$-space transformation $k_x\rightarrow -k_x+\pi$, it follows that the reduced zone scheme again contains two time-reversal invariant momenta at $k_x=0$ and $k_x=\pi/2$. As illustrated in Fig.~\ref{fig:psg81}, the reduced Brillouin zone harbors two degenerate sets of bands, the initial ones at $k_x\in[-\pi/2,\pi/2]$ and those originating from $k_x\in[\pi/2,3\pi/2]$. As an important difference compared to Sec.~\ref{sl81}, time reversal $\T$ comes along with an additional $k_x$-momentum shift of $\pi$, such that Kramer's pairs belong to {\it different} sets of bands. Accordingly, the modified $\mathds{Z}_2$ topological invariant from Eq.~(\ref{z2_invariant2}) needs to be adjusted,
\begin{equation}
\nu'=\frac{n'_{1\uparrow}-n'_{2\downarrow}}{2}\;\text{mod}\;2=\frac{n'_{2\uparrow}-n'_{1\downarrow}}{2}\;\text{mod}\;2\;,\label{z2_invariant3}
\end{equation}
where $n'_{1/2\uparrow/\downarrow}$ are Chern numbers evaluated in the reduced Brillouin zone. The first index $1/2$ refers to the set of bands ($1$ stands for the original bands at $k_x\in[-\pi/2,\pi/2]$ while $2$ denotes the shifted ones from $k_x\in[\pi/2,3\pi/3]$) and the second index $\uparrow/\downarrow$ specifies the spin sector. Given the Hamiltonian in Eqs.~(\ref{sl101_ham1}) and (\ref{sl101_ham2}), we find $\nu'=1$, demonstrating that a non-trivial topological index is established by time-reversal symmetry, coupling {\it different} sets of bands.

Labeling the Majorana modes at zero energy in the same way as in the previous section (i.e., by $\gamma_\text{R}$, $\gamma_\text{L}$, $\eta_\text{R}$, and $\eta_\text{L}$), we have calculated the symmetry properties of the coupling terms $i\gamma_\text{R}\gamma_\text{L}$, $i\eta_\text{R}\eta_\text{L}$, $i\gamma_\text{R}\eta_\text{L}$, and $i\gamma_\text{L}\eta_\text{R}$, see Table~\ref{psg101_table}. As indicated by the non-trivial $\mathds{Z}_2$ index $\nu'=1$, a coupling between the two sets of bands (via $i\gamma_\text{R}\eta_\text{L}$ or $i\gamma_\text{L}\eta_\text{R}$) is only possible by breaking time-reversal symmetry. Table~\ref{psg101_table} also shows that any linear combination of $i\gamma_\text{R}\eta_\text{L}$ and $i\gamma_\text{L}\eta_\text{R}$ additionally breaks translation invariance $T_x$. Even more, violating $\T$ and $T_x$ simultaneously is still not sufficient to gap out the edge states. A coupling between the two sets of bands can be achieved by a term $i\gamma_\text{R}\eta_\text{L}+i\gamma_\text{L}\eta_\text{R}$ (which also breaks $P_z$) or by a term $i\gamma_\text{R}\eta_\text{L}-i\gamma_\text{L}\eta_\text{R}$ (which also breaks $P_x$).

While the symmetry protection of edge states against a coupling between different momentum sectors turns out to be rather robust, the associated topological index $\nu'=1$ only remains valid if a coupling of boundary modes within the same set of bands is also forbidden by symmetry. Table~\ref{psg101_table} shows that this is indeed the case: There is no linear combination of $i\gamma_\text{R}\gamma_\text{L}$ and $i\eta_\text{R}\eta_\text{L}$ which satisfies all symmetries. Interestingly, since the term $i\gamma_\text{R}\gamma_\text{L}-i\eta_\text{R}\eta_\text{L}$ only breaks $P_x$ but no other symmetries, we identify reflection $P_x$ to be responsible for the protection. This also implies that for a cylinder without mirror symmetry (i.e., when the edges are along an arbitrary direction) the term $i\gamma_\text{R}\gamma_\text{L}-i\eta_\text{R}\eta_\text{L}$ is allowed by symmetries, turning the system into a trivial (spinon) superconductor.

In summary, we have investigated a spin phase where the boundary modes are protected by a modified $\mathds{Z}_2$ topological invariant. Edge states may only be gapped out by either breaking $\T$, $T_x$, and $P_z$ simultaneously, or by violating $P_x$. Hence, even though the mean-field ansatz is similar to the one in the previous section, the projective symmetries act very differently. Since the edge modes are entirely gapped out by only breaking $P_x$, the system is again similar to the previously introduced topological crystalline insulator.\cite{fu11}\\

%%%%%%%%%%%%%%%%%%%%%%%%%%%%%%%%%%%%%%%%%%%%%%%%%%%%%%%%%%%%%%
% V.F. EXAMPLE 6
%%%%%%%%%%%%%%%%%%%%%%%%%%%%%%%%%%%%%%%%%%%%%%%%%%%%%%%%%%%%%%

\subsection{PSG with $\{\eta_\T,\eta,\eta_1,\eta_2,\eta_z\}=\{1,1,-1,-1,1\}$ and $\{g_\T,g_{P_x},g_{P_y},g_{P_{xy}},g_{P_z}\}=\{i\tau^2,\tau^0,\tau^0,\tau^0,i\tau^3\}$}\label{sl628}
\begin{figure}[t]
\centering
\includegraphics[width=0.99\linewidth]{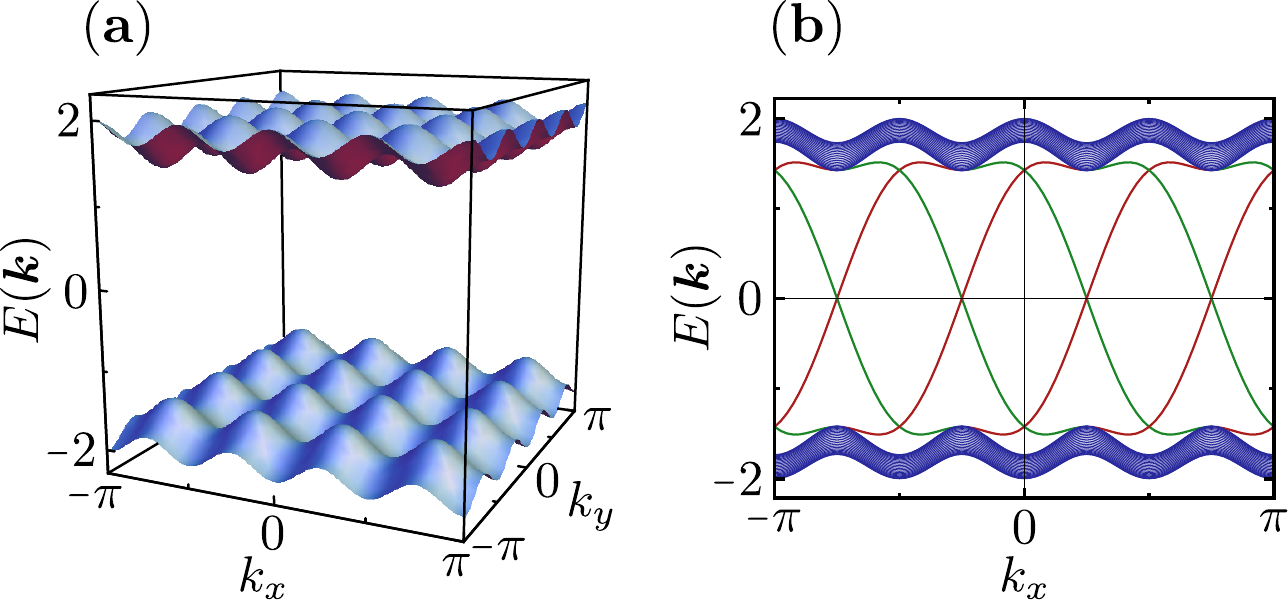}
\caption{Band structure in the PSG representation with $\{\eta_\T,\eta,\eta_1,\eta_2,\eta_z\}=\{1,1,-1,-1,1\}$ and $\{g_\T,g_{P_x},g_{P_y},g_{P_{xy}},g_{P_z}\}=\{i\tau^2,\tau^0,\tau^0,\tau^0,i\tau^3\}$ for a system (a) on a torus and (b) on a cylinder, using the parameters in Eq.~(\ref{sl628_parameters}). As illustrated in (b), the edge-state spectrum consists of four right-moving spin up modes (red) and four left-moving spin down modes (green). A detailed analysis with an enlarged unit cell (now comprising four sites in $x$-direction) shows that based on a modified $\mathds{Z}_2$ index, the band structure may be classified as topologically non-trivial.}
\label{fig:psg628}
\end{figure}

In a final example, we briefly illustrate that even more complex edge-state structures are possible. Considering the mean-field amplitudes
\begin{eqnarray}
&t_{2,\drr=(1,1)}^1=t_{2,\drr=(1,-1)}^1=-1\;,&\notag\\
&t_{3,\drr=(1,1)}^1=-t_{3,\drr=(1,-1)}^1=1\;,&\notag\\
&s_{\drr=(2,0)}^3=s_{\drr=(0,2)}^3=1\;,&\label{sl628_parameters}
\end{eqnarray}
the Hamiltonian reads as
\begin{equation}
H_\text{mf}=\frac{1}{2}\sum_\kk\hat{\Psi}_\kk^\dagger
\left(\begin{array}{cc}
h_\kk & 0 \\
0 & h^*_{-\kk}
\end{array}\right)
\hat{\Psi}_\kk
\end{equation}
with
\begin{widetext}
\begin{equation}
h_\kk=
\left(\begin{array}{cc}
\cos(2k_x)+\cos(2k_y) & 2i\sin(k_x)\cos(k_y)+2\cos(k_x)\sin(k_y)\\
-2i\sin(k_x)\cos(k_y)+2\cos(k_x)\sin(k_y) & -\cos(2k_x)-\cos(2k_y)
\end{array}\right)\;,
\end{equation}
\end{widetext}
where all Lagrange multipliers yet again vanish. Remarkably, the two spin sectors have Chern numbers $n_\uparrow=-n_\downarrow=4$ and the edge-state spectrum for the system on a cylinder exhibits {\it four} pairs of counter-propagating Majorana modes, see Fig.~\ref{fig:psg628}. Based on the trivial $\mathds{Z}_2$ index $\nu=0$, one might again expect the band structure to be topologically trivial. However, proceeding as above, it is again advantageous to describe the system by an enlarged unit cell, which now extends over four sites in  the $x$-direction. Within a reduced Brillouin zone $k_x\in[-\pi/4,\pi/4]$ one obtains four copies of degenerate sets of bands which are all characterized by Chern numbers $n'_\uparrow=-n'_\downarrow=1$ ($n'_\alpha$ is calculated in one quarter of the original Brillouin zone). In analogy to the previous sections, this allows us to define a modified $\mathds{Z}_2$ topological invariant $\nu'$. Since the system exhibits four right-moving ($\gamma_{1\text{R}},\ldots,\gamma_{4\text{R}}$) and four left-moving Majorana modes ($\gamma_{1\text{L}},\ldots,\gamma_{4\text{L}}$) at zero energy, there are in total 16 coupling terms $i\gamma_{a\text{R}}\gamma_{b\text{L}}$ with $a,b=\{1,2,3,4\}$. The index $\nu'$ only represents a topological invariant of the system if there is no linear combination of couplings $i\gamma_{a\text{R}}\gamma_{b\text{L}}$ that satisfies all symmetries. Given the large number of terms, we will not present a detailed symmetry analysis here, but conclude mentioning that the band structure is indeed topologically non-trivial, i.e., $\nu'=1$. Gapping out the edge states again require the breaking of various symmetries simultaneously. 
%
%%%%%%%%%%%%%%%%%%%%%%%%%%% Figure 7 %%%%%%%%%%%%%%%%%%%%%%%%%%%
%
%

%%%%%%%%%%%%%%%%%%%%%%%%%%%%%%%%%%%%%%%%%%%%%%%%%%%%%%%%%%%%%%
% VI. DISCUSSION
%%%%%%%%%%%%%%%%%%%%%%%%%%%%%%%%%%%%%%%%%%%%%%%%%%%%%%%%%%%%%%

\section{Discussion}\label{discussion}

The inception of topological electronic band insulators and superconductors highlighted the dramatic qualitative role that spin-orbit coupling can play in governing a material's physical properties.  As the field enters maturity, increasing attention is being focused on strongly correlated analogues.  Magnetic insulators that feature an interplay between strong spin-orbit coupling and frustration comprise prime candidates for the latter---particularly given the synthesis of new classes of compounds including the iridates that sharply violate SU(2) spin symmetry.  These systems can potentially host spin-liquid ground states where fractionalized spinon excitations (rather than electrons) exhibit topologically nontrivial band structures; a proof of concept has indeed been demonstrated in several earlier works, e.g., Refs.~\onlinecite{young08,rachel10,krempa10,pesin10,swingle11,ruegg12,cho12}. Compared to weakly correlated band insulators, however, our understanding of how spin-orbit coupling impacts spin-liquid physics remains in its infancy.   

In this work we have carried out a detailed PSG classification of $\mathds{Z}_2$ spin liquids on a square lattice when SU(2) spin symmetry is maximally broken.  These systems, while not of immediate experimental relevance, provide an ideal setting in which to explore the consequences of spin-orbit coupling on spin liquid phases since the SU(2)-invariant limit is so well-studied\cite{wen02,chen12, essin13}.  Within our PSG analysis we identify 1488 $\mathds{Z}_2$ spin liquids that {\it only} occur in the SU(2)-broken case and have no analogue among the 272 spin-isotropic $\mathds{Z}_2$ phases found in previous works. [Technically, as noted earlier eight of these 1488 states still preserve a projective version of SU(2) spin symmetry; see Ref.~\onlinecite{chen12}.]  Most strikingly, we have shown that these 1488 spin states naturally possess $p+ip$ pairing in the spinons leading in many cases to topologically non-trivial band structures associated with the existence of spinon edge states. Such spin phases hence realize the remarkable situation where a non-trivial topology occurs in two different contexts, primarily in the appearance of deconfined spinons as effective quasiparticles and secondarily through a topological spinon band structure.\cite{cho12,pesin10,ruegg12} The $\mathds{Z}_2$ gauge structure of our model Hamiltonians guarantees that gauge fluctuations emerging in a treatment beyond mean-field theory are fully gapped and should not qualitatively change the low-energy properties of the system.\cite{wen91}

We have discussed various examples of mean-field ans\"atze in selected PSG representations and investigated the origin of the topological edge-state protection. Our examples reveal a variety of different edge-state structures with up to four counter-propagating pairs of gapless Majorana modes. We expect that more involved mean-field ans\"atze may even possess higher numbers of boundary modes. Our analysis shows that many band structures can be characterized by a $\mathds{Z}_2$ topological index which, however, may differ from the $\mathds{Z}_2$ invariant for topological insulators. While our topological classification is built upon time-reversal symmetry, many PSG representations also require the fulfillment of additional lattice symmetries in order to retain the stability of boundary modes. Hence, we identify various spin liquids that may be referred to as ``topological crystalline spinon superconductors".  We have not attempted, however, to develop a unified picture characterizing the topology of all possible band structures.  A more involved task would be to clarify the role of $\mathds{Z}_2$ topological indices based \emph{solely} on lattice symmetries as considered in Refs.~\onlinecite{fu11,teo13,alexandradinata14}.  Other types of topological crystalline band structures could very well be lurking in our PSG analysis.  

%The examples described in Sec.~\ref{examples} are representative PSG states with spinon edge-state spectra that also occur in other spin-liquid solutions. The development of a unified picture characterizing the topology of all possible band structures, however, is beyond the scope of this work but would make for interesting future research.  While all $\mathds{Z}_2$ topological invariants identified in Sec.~\ref{examples} ultimately rely on time-reversal symmetry, a more involved task would be to clarify the role of $\mathds{Z}_2$ topological indices based on lattice symmetries as considered in Ref.~\onlinecite{fu11} and \onlinecite{teo13}. Possible ``topological crystalline insulator" spinon phases---in the sense of these works---could then also be identified among the PSG spin-liquid states.

It would be interesting to extend our analysis to other two-dimensional settings such as the Kagome lattice, where strong geometric frustration makes the existence of spin-liquid phases in generic spin Hamiltonians more likely.  We note that for the Kagome lattice a similar PSG spin-liquid classification has already been carried out in Ref.~\onlinecite{dodds13}. While this work considers the spin-symmetry breaking effects of Dzyaloshinskii-Moriya or Ising-spin terms, a residual U(1) spin symmetry is still preserved. In agreement with our results it is shown that the PSG classification itself remains unchanged when breaking down the spin symmetry to U(1); the spinon spectrum can, nevertheless, be significantly altered.  To the best of our knowledge, a complete PSG analysis in the case where the spin symmetry is maximally lifted has not been performed before.  

Another interesting extension of our work is the investigation of three-dimensional lattices.  Actually this setting promises to offer even richer physics upon including spin-orbit coupling since there one can have stable spin liquids with U(1) gauge freedom even if the spinons exhibit a bulk excitation gap (this is not so in two dimensions).  Thus spinon cousins of topological superconductors \emph{and} insulators\cite{pesin10,krempa10} can exist in three dimensions.  A spin liquid phase of the latter type has been proposed in the hyperkagome iridate compound Na$_4$Ir$_3$O$_8$.\cite{zhang08,takagi07}  In short, a systematic exploration of three-dimensional spin liquids with topological band structures induced by spin-orbit coupling should consider both U(1) and $\mathds{Z}_2$ states.  

Finally, while the model Hamiltonians of the present work are formulated in a fermionic parton representation for spins, it would be insightful to identify microscopic {\it spin} Hamiltonians that realize some of our phases as ground states. Such Hamiltonians may include spin-anisotropic Ising couplings $S_i^\mu S_j^\mu$, terms of Dzyaloshinskii-Moriya type $({\boldsymbol S}_i\times{\boldsymbol S}_j)^\mu$, or even contributions with more than two spin operators.  By backing out trial spin wavefunctions corresponding to a given spin-liquid ansatz, extensive variational studies could be employed to find such candidate Hamiltonians.  (At this point even mean-field energetics could be illuminating.)  Doing so will be a crucial step towards finding experimental candidates for these states, which is the ultimate goal of this study. 

\begin{acknowledgements}
The authors gratefully acknowledge illuminating conversations with Aris Alexandradinata, Andrew Essin, Roger Mong, and Frank Pollmann.  This research was supported by the Deutsche Akademie der Naturforscher Leopoldina through grant LPDS 2011-14 (J. R.); the NSF through grant DMR-1341822 (S.-P. L. and J. A.); the Alfred P. Sloan Foundation (J. A.); the Caltech Institute for Quantum Information and Matter, an NSF Physics Frontiers Center with support of the Gordon and Betty Moore Foundation; and the Walter Burke Institute for Theoretical Physics at Caltech.
\end{acknowledgements}

\appendix

%%%%%%%%%%%%%%%%%%%%%%%%%%%%%%%%%%%%%%%%%%%%%%%%%%%%%%%%%%%%%%
% APPENDIX A
%%%%%%%%%%%%%%%%%%%%%%%%%%%%%%%%%%%%%%%%%%%%%%%%%%%%%%%%%%%%%%

\section{Solutions for the matrices $g_\mathcal{S}$ in the spin-isotropic case}\label{g_solutions_1}
In this appendix, we show the 17 possible sets of matrices $g_\T$, $g_{P_x}$, $g_{P_y}$, $g_{P_{xy}}$ defining the PSG gauge transformations $G^\mathcal{S}_\rr=d^\mathcal{S}_\rr g_\mathcal{S}$ when SU(2) spin symmetry is intact. These solutions follow from Eq.~(\ref{gggg}) and are given by
\begin{equation}
g_\T=\tau^0\;,\; g_{P_x}=\tau^0\;,\; g_{P_y}=\tau^0\;,\; g_{P_{xy}}=\tau^0\;;\label{su2_first}
\end{equation}
\begin{equation}
g_\T=\tau^0\;,\; g_{P_x}=i\tau^3\;,\; g_{P_y}=i\tau^3\;,\; g_{P_{xy}}=\tau^0\;;\label{case_2}
\end{equation}
\begin{equation}
g_\T=\tau^0\;,\; g_{P_x}=\tau^0\;,\; g_{P_y}=\tau^0\;,\; g_{P_{xy}}=i\tau^3\;;\label{case_3}
\end{equation}
\begin{equation}
g_\T=\tau^0\;,\; g_{P_x}=i\tau^3\;,\; g_{P_y}=i\tau^3\;,\; g_{P_{xy}}=i\tau^3\;;\label{case_4}
\end{equation}
\begin{equation}
g_\T=\tau^0\;,\; g_{P_x}=i\tau^1\;,\; g_{P_y}=i\tau^1\;,\; g_{P_{xy}}=i\tau^3\;;\label{case_5}
\end{equation}
\begin{equation}
g_\T=i\tau^2\;,\; g_{P_x}=\tau^0\;,\; g_{P_y}=\tau^0\;,\; g_{P_{xy}}=\tau^0\;;\label{case_6}
\end{equation}
\begin{equation}
g_\T=i\tau^2\;,\; g_{P_x}=i\tau^2\;,\; g_{P_y}=i\tau^2\;,\; g_{P_{xy}}=\tau^0\;;\label{case_7}
\end{equation}
\begin{equation}
g_\T=i\tau^2\;,\; g_{P_x}=i\tau^3\;,\; g_{P_y}=i\tau^3\;,\; g_{P_{xy}}=\tau^0\;;\label{case_8}
\end{equation}
\begin{equation}
g_\T=i\tau^2\;,\; g_{P_x}=\tau^0\;,\; g_{P_y}=\tau^0\;,\; g_{P_{xy}}=i\tau^2\;;\label{case_9}
\end{equation}
\begin{equation}
g_\T=i\tau^2\;,\; g_{P_x}=i\tau^2\;,\; g_{P_y}=i\tau^2\;,\; g_{P_{xy}}=i\tau^2\;;\label{case_10}
\end{equation}
\begin{equation}
g_\T=i\tau^2\;,\; g_{P_x}=i\tau^3\;,\; g_{P_y}=i\tau^3\;,\; g_{P_{xy}}=i\tau^2\;;\label{case_11}
\end{equation}
\begin{equation}
g_\T=i\tau^2\;,\; g_{P_x}=\tau^0\;,\; g_{P_y}=\tau^0\;,\; g_{P_{xy}}=i\tau^3\;;\label{case_12}
\end{equation}
\begin{equation}
g_\T=i\tau^2\;,\; g_{P_x}=i\tau^2\;,\; g_{P_y}=i\tau^2\;,\; g_{P_{xy}}=i\tau^3\;;\label{case_13}
\end{equation}
\begin{equation}
g_\T=i\tau^2\;,\; g_{P_x}=i\tau^3\;,\; g_{P_y}=i\tau^3\;,\; g_{P_{xy}}=i\tau^3\;;\label{case_14}
\end{equation}
\begin{equation}
g_\T=i\tau^2\;,\; g_{P_x}=i\tau^3\;,\; g_{P_y}=i\tau^3\;,\; g_{P_{xy}}=i\tau^1\;;\label{case_15}
\end{equation}
\begin{equation}
g_\T=\tau^0\;,\; g_{P_x}=i\tau^1\;,\; g_{P_y}=i\tau^2\;,\; g_{P_{xy}}=\frac{i\tau^1+i\tau^2}{\sqrt{2}}\;;\label{case_16}
\end{equation}
\begin{equation}
g_\T=i\tau^2\;,\; g_{P_x}=i\tau^1\;,\; g_{P_y}=i\tau^3\;,\; g_{P_{xy}}=\frac{i\tau^1+i\tau^3}{\sqrt{2}}\;.\label{su2_last}
\end{equation}
Note that in some of these equations we have used a different gauge as compared to Ref.~\onlinecite{wen02}, such that $g_\T$ is either $g_\T=\tau^0$ or $g_T=i\tau^2$. This gauge choice is convenient for our discussion in Sec.~\ref{examples}, because $g_\T=i\tau^2$ in conjunction with $\eta_\T=1$ implies that projective time-reversal symmetry and the time-reversal symmetry of an ordinary fermion system are identical.

%%%%%%%%%%%%%%%%%%%%%%%%%%%%%%%%%%%%%%%%%%%%%%%%%%%%%%%%%%%%%%
% APPENDIX B
%%%%%%%%%%%%%%%%%%%%%%%%%%%%%%%%%%%%%%%%%%%%%%%%%%%%%%%%%%%%%%

\section{Solutions for the matrices $g_\mathcal{S}$ when SU(2) spin symmetry is maximally lifted}\label{g_solutions_2}
Here we show all solutions of Eqs.~(\ref{gggg}) and (\ref{gggg2}) resulting in sets of matrices $\{g_\T,g_{P_x},g_{P_y},g_{P_{xy}},g_{P_z}\}$. The 17 solutions for $g_\T$, $g_{P_x}$, $g_{P_y}$, $g_{P_{xy}}$ obtained from Eq.~(\ref{gggg}) have already been listed in Appendix~\ref{g_solutions_1}. Hence, we only need to show the possible solutions for $g_{P_z}$ in each of these cases. In total, there are 55 different sets of matrices given by
\begin{equation}
\text{(\ref{su2_first})}\;\Longrightarrow\; g_{P_z}=\tau^0\;\text{or}\quad i\tau^3\;,
\end{equation}
\begin{equation}
\text{(\ref{case_2})}\;\Longrightarrow\; g_{P_z}=\tau^0\;\text{or}\quad i\tau^2\;\text{or}\quad i\tau^3\;,
\end{equation}
\begin{equation}
\text{(\ref{case_3})}\;\Longrightarrow\; g_{P_z}=\tau^0\;\text{or}\quad i\tau^2\;\text{or}\quad i\tau^3\;,
\end{equation}
\begin{equation}
\text{(\ref{case_4})}\;\Longrightarrow\; g_{P_z}=\tau^0\;\text{or}\quad i\tau^2\;\text{or}\quad i\tau^3\;,
\end{equation}
\begin{equation}
\text{(\ref{case_5})}\;\Longrightarrow\; g_{P_z}=\tau^0\;\text{or}\quad i\tau^1\;\text{or}\quad i\tau^2\;\text{or}\quad i\tau^3\;,
\end{equation}
\begin{equation}
\text{(\ref{case_6})}\;\Longrightarrow\; g_{P_z}=\tau^0\;\text{or}\quad i\tau^2\;\text{or}\quad i\tau^3\;,
\end{equation}
\begin{equation}
\text{(\ref{case_7})}\;\Longrightarrow\; g_{P_z}=\tau^0\;\text{or}\quad i\tau^2\;\text{or}\quad i\tau^3\;,
\end{equation}
\begin{equation}
\text{(\ref{case_8})}\;\Longrightarrow\; g_{P_z}=\tau^0\;\text{or}\quad i\tau^1\;\text{or}\quad i\tau^2\;\text{or}\quad i\tau^3\;,
\end{equation}
\begin{equation}
\text{(\ref{case_9})}\;\Longrightarrow\; g_{P_z}=\tau^0\;\text{or}\quad i\tau^2\;\text{or}\quad i\tau^3\;,
\end{equation}
\begin{equation}
\text{(\ref{case_10})}\;\Longrightarrow\; g_{P_z}=\tau^0\;\text{or}\quad i\tau^2\;\text{or}\quad i\tau^3\;,
\end{equation}
\begin{equation}
\text{(\ref{case_11})}\;\Longrightarrow\; g_{P_z}=\tau^0\;\text{or}\quad i\tau^1\;\text{or}\quad i\tau^2\;\text{or}\quad i\tau^3\;,
\end{equation}
\begin{equation}
\text{(\ref{case_12})}\;\Longrightarrow\; g_{P_z}=\tau^0\;\text{or}\quad i\tau^1\;\text{or}\quad i\tau^2\;\text{or}\quad i\tau^3\;,
\end{equation}
\begin{equation}
\text{(\ref{case_13})}\;\Longrightarrow\; g_{P_z}=\tau^0\;\text{or}\quad i\tau^1\;\text{or}\quad i\tau^2\;\text{or}\quad i\tau^3\;,
\end{equation}
\begin{equation}
\text{(\ref{case_14})}\;\Longrightarrow\; g_{P_z}=\tau^0\;\text{or}\quad i\tau^1\;\text{or}\quad i\tau^2\;\text{or}\quad i\tau^3\;,
\end{equation}
\begin{equation}
\text{(\ref{case_15})}\;\Longrightarrow\; g_{P_z}=\tau^0\;\text{or}\quad i\tau^1\;\text{or}\quad i\tau^2\;\text{or}\quad i\tau^3\;,
\end{equation}
\begin{equation}
\text{(\ref{case_16})}\;\Longrightarrow\; g_{P_z}=\tau^0\;\text{or}\quad i\tau^3\;,
\end{equation}
\begin{equation}
\text{(\ref{su2_last})}\;\Longrightarrow\; g_{P_z}=\tau^0\;\text{or}\quad i\tau^2\;.
\end{equation}

%\bibliographystyle{prsty}
%\bibliography{refs}

\begin{thebibliography}{10}
\bibitem{pxy}
Note that the symmetries $P_x$, $P_y$, and $P_{xy}$ are not independent of each
  other, since they fulfill the relation $P_y=P_{xy}P_x P_{xy}$. Alternatively,
  one could apply the PSG method with dependent symmetries by using $90^\circ$
  rotation symmetry ($P_{\pi/2}$) instead of $P_{xy}$. However, it turns out to
  be convenient to carry out the classification procedure with point group
  symmetries that square to one. Working with a set of dependent symmetries has
  no consequences on our analysis.
\bibitem{t}
In this gauge $\T^2$ acts as $\T^2|\phi\rangle=|\phi\rangle$. This shows that
  within a projective treatment, the sign of $\T^2$ is not uniquely determined,
  but depends on the specific gauge.
\bibitem{almost_all}
Exceptions are Eqs.~(\ref{case_16}) and (\ref{su2_last}) where $g_{P_{xy}}$ is
  given by linear combinations of two Pauli matrices. While the mean-field
  ans{\"a}tze in these PSG representations have a more complicated structure
  which we are not going to further elaborate on, their spinon dispersions do
  not qualitatively differ from those of other PSG representations.
\bibitem{lagrange}
Vanishing Lagrange multipliers are often a consequence of our simple choices
  for the mean-field parameters, neglecting longer range hoppings or pairings.
\bibitem{z2_index}
To show that the $\mathds{Z}_2$ classification remains valid when coupling the
  two sets of bands we have only considered boundary modes at $k_x=0$. In
  principle, the same analysis also needs to be performed for possible
  edge-state crossings at $k_x=\frac{\pi}{2}$, which might occur upon deforming
  the bands. However, we could not identify any parameter regime in this PSG
  phase with such a level crossing. Furthermore, given the symmetry properties
  of the Majorana modes at $k_x=0$ we find that the momenta $k_x=0$ and
  $k_x=\frac{\pi}{2}$ generally exhibit the same edge-state protection.
\end{thebibliography}

\end{document}